\documentclass[journal]{IEEEtran}
\usepackage{graphicx}
 \usepackage{algorithm}
\usepackage{algorithmic}
 \usepackage{subfigure}
\usepackage{amsmath,amssymb,amsfonts}

 \usepackage{url}
 \usepackage{amsmath,bm}
 \usepackage{url}
\usepackage{multirow}
\usepackage{makecell}
\usepackage{cite}

\ifCLASSINFOpdf

\else

\fi

\begin{document}

\title{A VHetNet-Enabled Asynchronous Federated Learning-Based Anomaly Detection Framework for Ubiquitous IoT}

\author{Weili~Wang,
         Omid~Abbasi,~\IEEEmembership{Senior Member,~IEEE},
         Halim~Yanikomeroglu,~\IEEEmembership{Fellow,~IEEE},
        Chengchao~Liang,
        Lun~Tang,
         Qianbin~Chen,~\IEEEmembership{Senior Member,~IEEE}
\thanks{This work was supported by National Natural Science Foundation of China (Grant No. 62071078 and Grant No. 62001076). \emph{(Corresponding author: Qianbin Chen.)}}
\thanks{W. Wang is with the School of Communication and Information Engineering and the Key Laboratory of Mobile Communication,
Chongqing University of Posts and Telecommunications, Chongqing 400065, China, and also with the Department of Systems and
Computer Engineering, Carleton University, Ottawa, ON K1S5B6, Canada (e-mail: 1961797154@qq.com).}
\thanks{O. Abbasi and H. Yanikomeroglu are with the Department of Systems and
Computer Engineering, Carleton University, Ottawa, ON K1S5B6, Canada (e-mail: omidabbasi@sce.carleton.ca; halim@sce.carleton.ca).}
\thanks{C. Liang, L. Tang and Q. Chen are with the School of Communication and Information Engineering and the Key Laboratory of Mobile Communication,
Chongqing University of Posts and Telecommunications, Chongqing
400065, China (e-mail: liangcc@cqupt.edu.cn; tangl@cqupt.edu.cn; cqb@cqupt.edu.cn).}}

\maketitle

\begin{abstract}
Anomaly detection for the Internet of Things (IoT) is a major intelligent service required by many fields, including intrusion detection, state monitoring, device-activity analysis, and security supervision. However, the heterogeneous distribution of data and resource-constrained end nodes in ubiquitous IoT systems present challenges for existing anomaly detection models. Due to the advantages of flexible deployment and multi-dimensional resources, high altitude platform stations (HAPSs) and unmanned aerial vehicles (UAVs), which are important components of vertical heterogeneous networks (VHetNets), have significant potential for sensing, computing, storage, and communication applications in ubiquitous IoT systems. In this paper, we propose a novel VHetNet-enabled asynchronous federated learning (AFL) framework to enable decentralized UAVs to collaboratively train a global anomaly detection model based on their local sensory data from ubiquitous IoT devices. In the VHetNet-enabled AFL framework, a HAPS operates as a central aerial server, and the local models trained in UAVs are uploaded to the HAPS for global aggregation due to its wide coverage and strong storage and computation capabilities. We also introduce a UAV selection strategy into the AFL framework to prevent UAVs with low local model quality and large energy consumption from affecting the learning efficiency and detection accuracy of the global model. To ensure the security of transmissions between UAVs and the HAPS via wireless links, we add designed noise to local model parameters in UAVs to achieve differential privacy during the information exchange process. Moreover, we propose a compound-action actor-critic (CA2C)--based joint device association, UAV selection, and UAV trajectory planning algorithm to further enhance the overall federated execution efficiency and detection model accuracy under the UAV energy constraints. Extensive experimental evaluation on a real-world dataset demonstrates that the proposed algorithm can achieve high detection accuracy with short federated execution time and low energy consumption.
\end{abstract}

\begin{IEEEkeywords}
Anomaly detection, ubiquitous Internet of Things (IoT), vertical heterogeneous network (VHetNet), asynchronous federated learning (AFL), differential privacy.
\end{IEEEkeywords}

\IEEEpeerreviewmaketitle

\section{Introduction}
\bstctlcite{IEEEexample:BSTcontrol}
\IEEEPARstart{C}{ompared} to existing wireless technologies including 5G, 6G and beyond networks represent more than an improvement of basic performance requirements, such as increased data rates, reduced latency, and enhanced spectrum and energy efficiency \cite{2-9749222}. 6G and beyond networks are envisioned to have some unique characteristics. First, space networks (including satellites in low, medium and geostationary Earth orbits), air networks (including high altitude platform stations (HAPSs) and unmanned aerial vehicles (UAVs)), and terrestrial networks (including macro and micro base stations), are integrated into vertical heterogeneous networks (VHetNets) to provide global coverage \cite{3-9749175,5-9535454}. Second, numerous devices, such as environmental monitoring sensors, healthcare wearables, and industrial control agents, are joining Internet of Things (IoT) networks. The ubiquitous IoT is expected to support seamless connectivity anytime, anywhere, and for everything \cite{4-9606808}. Third, artificial intelligence (AI) penetrates into every corner of networks, ranging from end devices to the core. Network nodes are being endowed with built-in AI, which will support diversified AI services and facilitate intelligent network management \cite{6-8808168}.

Anomaly detection is defined as a process of automatically detecting whether devices, components, or systems are running normally or not \cite{7-9301388}. Anomaly detection is one of the indispensable AI services required by ubiquitous IoT devices for early warnings of abnormal behaviors and uninterrupted operations. Data is the basis of anomaly detection, and machine learning (ML) is the most commonly used technique for anomaly detection models \cite{17-9116088}. Traditional anomaly detection schemes in IoT usually utilize static sensors to perform data sensing and IoT gateways to forward the sensory data to a central server to learn a global anomaly detection model \cite{10-8633365,9-8896029,8-8986829}. However, in situations where sensing targets are continually moving or are located in far-flung regions, traditional sensing solutions will experience significant challenges. In addition, due to the explosion of data in ubiquitous IoT, privacy concern and limited communication resources for data transmission present a major impediment to any centralized learning framework.

To implement anomaly detection in ubiquitous IoT, HAPSs and UAVs represent advanced approaches for smart sensing and data analysis \cite{1-9453811,11-9543534,12-9444660,13-9130055,14-9462712}. A three-layer VHetNet consisting of a HAPS, UAVs, and IoT devices is a promising architecture for learning a global anomaly detection model, where UAVs are deployed as flying sensors for data sensing from IoT devices, and the HAPS is deployed as a central aerial server for data analysis and network control. Compared to static sensors, using UAVs as aerial nodes to provide wireless sensing support is a promising paradigm due to their wider field of view, highly flexible and controllable 3D mobility, and the capability of sensing performance optimization through UAV trajectory planning \cite{15-9456851}. In addition, a HAPS can work as an efficient central aerial server due to its quasi-stationary position, line-of-sight communication, wide coverage, and multiple energy sources, including conventional energy (such as electrical batteries and fuel tanks), energy beams, and solar energy \cite{16-9380673}.

As an emerging decentralized learning framework, federated learning (FL) has been shown to be effective in overcoming the challenges of privacy concerns and limited communication resources by enabling distributed UAVs to train a shared anomaly detection model locally without data centralization \cite{4-9606808}. By using FL, a UAV can perform local model training based on sensory data from its associated IoT devices, and local models from all UAVs are uploaded to the HAPS periodically for global aggregation. In general, compared to a centralized training framework, FL allows the VHetNet to learn a global anomaly detection model in a secure and efficient way. However, since there must be a periodic exchange of all UAV model parameters with the HAPS through wireless links during the model training process, the VHetNet-enabled FL framework still faces some challenges.
\begin{itemize}
\item The FL convergence will inevitably be affected by the learning latency and model quality of each UAV.
\item Because UAVs and HAPSs operate in the sky and communicate through wireless communication technologies, their sensitive information is more likely to be inferred by the shared parameters between them.
\item UAVs are generally energy-constrained, and hence balancing energy usage between tracking moving targets, computation, and transmitting data for model training is a thorny problem.
\end{itemize}

These challenges pertaining to anomaly detection and FL require an in-depth investigation into the efficient implementation of anomaly detection in the VHetNet-enabled FL framework.

\subsection{Related Work}
In recent years, research on FL-based distributed learning frameworks has received much attention in wireless networks \cite{18-9237167,19-9485089,20-9062302,21-9528995,22-9127160,23-8998397}, where the main role of FL has been to preserve data privacy and improve learning efficiency. Liu \emph{et al.} \cite{18-9237167}, for instance, formulated an FL reinforcement learning framework that relied on radio access network slicing to achieve an efficient device association scheme, where FL was adopted to promote the collaboration between smart devices while reducing the bandwidth consumption for learning. In \cite{19-9485089}, FL was used with edge intelligence--powered small-cell networks to decrease the signaling overhead and computing complexity of base stations when collaborating to learn an intelligent computation offloading and interference coordination scheme. Wang \emph{et al.} \cite{20-9062302} proposed an FL-based cooperative edge caching framework to enable all local users to cooperatively learn a shared content popularity prediction model, which could accelerate the convergence speed and improve the hit rate. However, in \cite{18-9237167,19-9485089,20-9062302}, all local agents were required to participate in each global aggregation, which neglected the influence of poor model quality and slow model updates of some local agents on learning efficiency and global model accuracy. To solve this problem, some studies have used a device selection scheme for FL \cite{21-9528995,22-9127160,23-8998397}. To balance the tradeoff between model accuracy and FL costs in wireless networks, Zhao \emph{et al.} \cite{21-9528995} formulated a joint resource allocation, data management, and user selection optimization problem and solved it with a computation-efficient algorithm. In \cite{22-9127160}, the authors proposed a hierarchical federated edge learning framework and an efficient edge association and resource scheduling algorithm to minimize both the system energy and FL execution time. A deep reinforcement learning (DRL)--based node selection algorithm was developed in \cite{23-8998397} to improve efficiency and training accuracy in an asynchronous federated learning (AFL) framework. Theoretically, AFL frameworks have great potential to establish a global ML model with improved learning efficiency and model accuracy in a distributed IoT environment.

Using ML models to detect abnormal behaviors among IoT devices has attracted many research efforts with a view to securing critical infrastructure \cite{24-8430652,8-8986829,26-9261961,27-9146846,28-9522027}. In \cite{8-8986829}, the authors designed an integrated model involving a convolutional neural network with long-short term memory for anomaly detection in an IoT time series. A one-class support Tucker machine method was designed in \cite{24-8430652} for unsupervised outlier detection of IoT big data.  However, the framework used in \cite{8-8986829} and \cite{24-8430652} relied on a central server to detect anomalies, which would cause network congestion and computing pressure for the central server. Some studies have proposed a distributed framework approach to address these challenges, where FL is the prevailing distributed technique \cite{26-9261961,27-9146846,28-9522027}. In \cite{26-9261961}, the authors used clustering and classification methods to build a novel anomaly detection framework for hypertext transfer protocol, which drew on edge intelligence for IoT to distribute the entire detection process. To secure manufacturing processes for the industrial IoT, an FL-based distributed learning framework was proposed in \cite{27-9146846} to combine all edge devices to train a global anomaly detection model. To address the efficiency, robustness, and security challenges facing FL, Cui \emph{et al.} \cite{28-9522027} proposed a differentially private AFL-based anomaly detection scheme for IoT infrastructure. However, in the literature on anomaly detection methods for IoT environments, there is a lack of discussion about the issue of continually moving sensing targets. Yet this is an issue that increases the difficulty of data sensing and collection.

UAVs and HAPSs can provide wireless sensing support from the sky to track moving sensing targets \cite{32-9154432,33-9374461,14-9462712}. In \cite{32-9154432}, a cooperative internet of UAVs was established to execute integrated sensing and transmission tasks for ground sensing targets under energy and communication constraints. In \cite{33-9374461}, the authors proposed a joint sensing and transmission protocol to enable UAV-to-Device communication to improve UAV sensing services. Kurt \emph{et al.} \cite{14-9462712} leveraged caching, communication, computing, and sensing capabilities of a HAPS to serve autonomous devices for future aerial delivery networks. In addition to providing sensing capabilities, UAV-enabled and HAPS-enabled wireless technologies also play key roles in enhancing wireless connectivity and computation capability for ubiquitous IoT \cite{29-9552467,30-9149835,31-8432464,34-9772280}. Considering the constrained energy capacity in UAVs, \cite{29-9552467,30-9149835,31-8432464} studied the UAV-enabled energy-efficient computation offloading, data uploading, and coverage enhancement systems, respectively. Compared to UAVs, HAPSs can provide wider coverage and stronger computational capabilities in a sustainable manner. Ren \emph{et al.} \cite{34-9772280} proposed a HAPS-assisted caching and computation offloading framework for intelligent transportation systems, where a HAPS played the dual role of a powerful computing server and a data library. Nevertheless, how to utilize the limited onboard energy storage in UAVs to complete different network tasks is still a vexing problem.

The secure information transmission between IoT devices and aerial networks is necessary to ensure data privacy and security. In \cite{50-9615108}, a secure identity-free transmission strategy was proposed to support UAV-aided IoT networks, where IoT devices uploaded their data packets to UAVs without location and identity information. Tang \emph{et al.} \cite{51-9915381} proposed a UAV-assisted data collection scheme for clustered IoT devices with a proof-of-stake blockchain-based security technique. In a VHetNet-enabled FL framework, a security strategy is needed to avoid parameter inference attacks. In \cite{11-9543534}, a privacy-preserving AFL framework for multi-UAV-enabled networks was developed for distributed ML model training, and a multi-agent asynchronous advantage actor-critic-based resource scheduling algorithm was proposed to minimize the federated execution time and learning accuracy loss. However, to the best of our knowledge, there are no studies that have considered anomaly detection and network scheduling jointly in a VHetNet-enabled FL framework, which is of great importance to achieve intelligent network management for 6G and beyond networks.


\subsection{Contributions and Organization}
To tackle the aforementioned challenges, we propose a VHetNet-enabled AFL framework to implement the self-scheduling
anomaly detection model, which can achieve high learning efficiency and model accuracy with the assistance of a network scheduling strategy. The proposed framework enables UAVs to train the anomaly detection model locally and asynchronously upload model parameters to a HAPS for global aggregation without raw data transmission. This reduces the communication and computation cost as well as improves the learning efficiency and accuracy. To ensure the security of the transmitted information, we add designed noise to local model parameters to achieve differential privacy during the information exchange process. Considering the limited onboard energy storage and distinct model quality of UAVs, we also propose a compound-action actor-critic (CA2C)--based joint device association, UAV selection, and UAV trajectory planning algorithm to facilitate the efficient implementation of the self-scheduling anomaly detection model. More specifically, the main contributions of this paper are as follows:
\begin{itemize}
\item By using IoT devices to provide data support, UAVs to sense data from IoT devices and local model training, and a HAPS to execute the control agent and global aggregation, we develop a VHetNet-enabled AFL framework to provide an anomaly detection service for ubiquitous IoT. In the proposed framework, UAVs with high local model quality and low energy consumption are selected to participate in the global aggregation instead of waiting for all UAVs to complete their local model update.
\item A differentially private Wasserstein generative adversarial network with gradient penalty (WGAN-GP) is proposed as the basic anomaly detection model, where differential privacy is achieved by adding carefully designed noise to gradients during the learning process.
\item A joint device association, UAV selection, and UAV trajectory planning problem is formulated and modeled as a dynamic non-convex combination problem to maximize the network coverage capacity and minimize the overall federated execution time and learning accuracy loss. In the context of the dynamic VHetNet environment, we introduce a CA2C-based self-scheduling solution to determine both the discrete (including device association and UAV selection) and continuous (including UAV trajectory planning) actions dynamically.
\item We conduct extensive experimental evaluation on a real-world dataset to validate the efficiency and effectiveness of the proposed CA2C-AFL-based anomaly detection algorithm. Simulation results demonstrate that the proposed algorithm can achieve high detection accuracy with short federated execution time and low energy consumption.
\end{itemize}

The rest of the paper is organized as follows. Section II presents the system model and problem formulation of the VHetNet-enabled AFL-based anomaly detection framework. The CA2C-based self-scheduling solution is proposed in Section III to facilitate the implementation of the formulated framework. The performance of the proposed CA2C-AFL based anomaly detection algorithm is evaluated in Section IV, and Section V concludes the paper.

\section{System Model and Problem Formulation}
Fig. 1 shows a three-layer VHetNet-enabled AFL framework, which consists of a HAPS, $N$ UAVs, and $K$ single-antenna IoT devices, denoted by $ \mathcal{N} = \{ 1,...,N\}$ and $\mathcal{K} = \{ 1,...,K\}$, respectively. In the proposed framework, the main roles of each layer are elaborated as follows:

\begin{figure}[!t]
\centering
\includegraphics[width=3.5in]{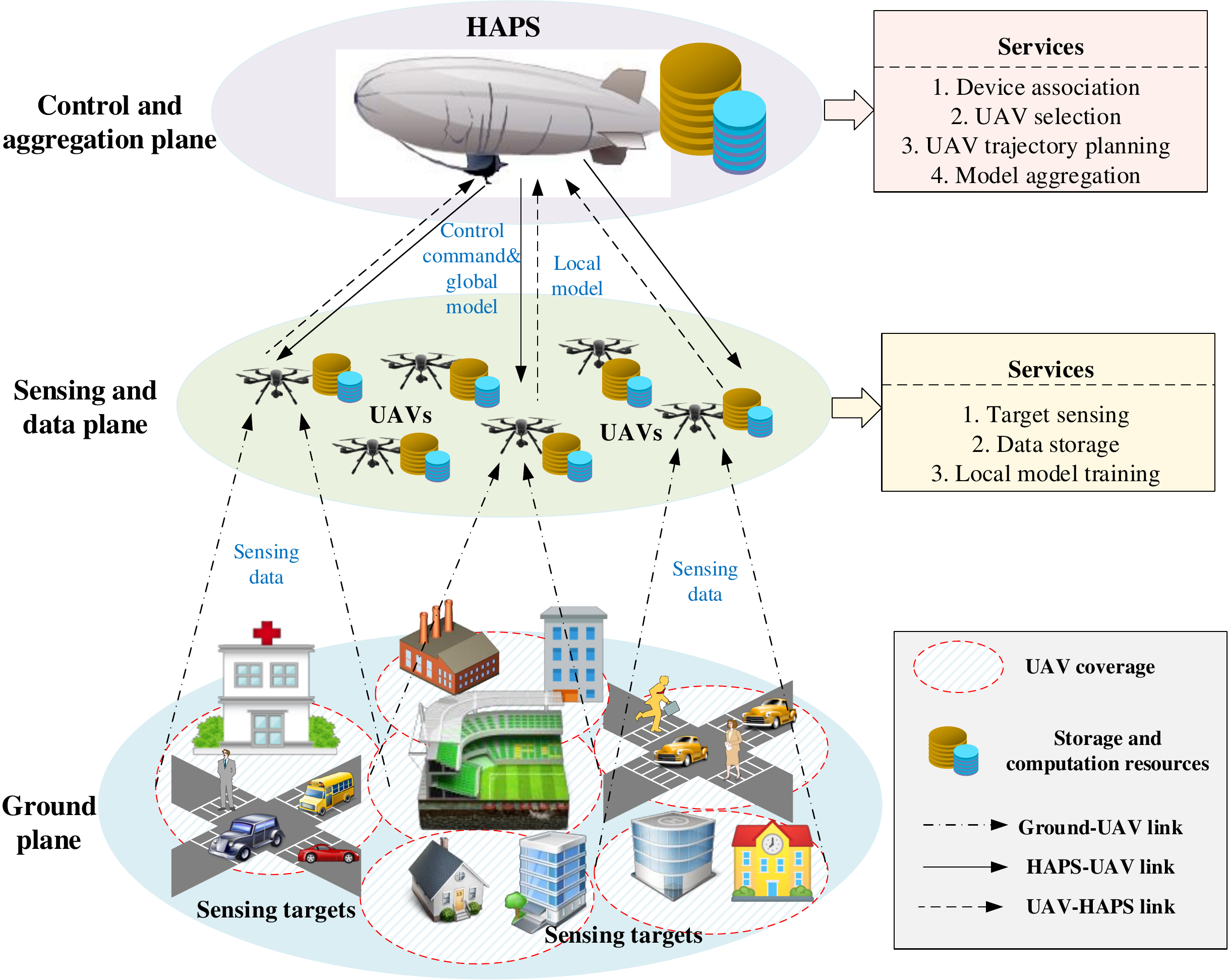}
\caption{A three-layer VHetNet-enabled AFL framework.}
\end{figure}

\textbf{Ground plane}. The ground plane consists of various IoT devices, such as environmental monitoring sensors, healthcare wearables, and industrial control agents. Multiple IoT devices will produce large amounts of heterogeneous data, which can provide data
support for the implementation of the anomaly detection model. By collecting and analyzing the sensory data indicating the running states of different systems, the anomaly detection model can be constructed to detect abnormal behaviors in monitored systems and provide early warnings before the occurrence of failure.

\textbf{Sensing and data plane}. This plane is composed of  UAVs, and each UAV plays the role of a sensor, computing node, and storage node. A UAV is responsible for sensing the ground IoT devices within its coverage area and training the local learning model using its sensory data. Using UAVs as aerial nodes to provide wireless sensing support from the sky is a promising paradigm for three reasons: the wider field of view, the highly flexible and controllable 3D mobility, and the capability of sensing performance optimization through UAV trajectory planning \cite{15-9456851}.

\textbf{Control and aggregation plane}. A HAPS in the control and aggregation plane works as the aerial center server. As a quasi-stationary network node, a HAPS operates in the stratosphere at an altitude of around 20 km \cite{16-9380673}. Due to the advantages of its wide coverage area and strong computational capabilities, a HAPS can control and manage the whole network intelligently using AI techniques. In the proposed framework, the HAPS is responsible for network control and the global aggregation of local models from UAVs.

\subsection{UAV Sensing}
In our proposed VHetNet-enabled AFL framework, each UAV needs to sense data from its associated IoT devices and store the sensory data to update its local learning model periodically. A UAV is required to support the sensing and the local model update for $T$ time slots with its constrained energy. At time slot $t \in \mathcal{T} (\mathcal{T} = \{ 1,...,T\} )$, we denote the location of device $k \in \mathcal{K}$ and the location of UAV $n \in \mathcal{N}$ by $\bm{x}_k (t) = (x_k (t),y_k (t),0)$ and $\bm{x}_n (t) = (x_n (t),y_n (t),z_n (t))$, respectively. In addition, we assume that the altitude $z_n (t)$ of each UAV $n$ does not change during any $T$  time slots, such that $z_n (t) \triangleq H,{\text{ }}\forall n,t$. Hence, the distance between the UAV $n$ and device $k$ is calculated by
\begin{equation}
d_{n,k} (t) = \sqrt {(x_n (t) - x_k (t))^2  + (y_n (t) - y_k (t))^2  + H^2 }.
\end{equation}

We introduce the expression $\lambda _{n,k} (t)$ to denote whether or not device $k$ is associated with  UAV $n$, with $\lambda _{n,k} (t) = 1$ indicating that device $k$ is associated with  UAV $n$, and $\lambda _{n,k} (t) = 0$ indicating otherwise.  If $\lambda _{n,k} (t) = 1$, then UAV $n$ is responsible for the sensing of device $k$ at time slot $t$. The successful sensing probability for an IoT device can be expressed as an exponential function of the distance between the UAV $n$ and the device $k$ \cite{13-9130055,33-9374461}, which is defined as
\begin{equation}
P_{n,k} (t) = \lambda _{n,k} (t)e^{ - \xi d_{n,k} (t)},
\end{equation}
where $\xi$ is a parameter reflecting the sensing performance. To express the coverage capacity of UAVs, we set a minimum successful sensing probability threshold $P_{{\text{th}}}$ for UAVs. When UAV $n$ senses the device $k$ for the real-time monitoring at time slot $t$, the successful sensing probability should satisfy
\begin{equation}
P_{n,k} (t) \geqslant P_{{\text{th}}} ,\;\forall k, n, t.
\end{equation}

Hence, the coverage capacity of UAV $n$ at time slot $t$ is expressed as the total number of successfully sensed devices, which is given by
\begin{equation}
C_n^C (t) = \sum\limits_{k = 1}^K {1_{\{ P_{n,k} (t) \geqslant P_{{\text{th}}} \} } }.
\end{equation}

\subsection{AFL-Based Anomaly Detection Model}
FL enables UAVs to execute local training using their own sensory data and avoid raw data transmission to the HAPS for privacy preservation and communication efficiency. Traditional FL commonly uses a synchronous learning mechanism to aggregate local models from all distributed agents \cite{37-konecny2017federated}. Since the local model quality and battery power of different UAVs will differ during operations, the FL framework, which needs to wait for all UAVs to finish their local training before the global aggregation, will inevitably result in unstable model quality and increased learning latency. To address this, we develop a VHetNet-enabled AFL-based anomaly detection framework for ubiquitous IoT. Meanwhile, the UAV selection strategy is introduced into the AFL framework to keep UAVs with low local model quality and large energy consumption from affecting the learning efficiency and model accuracy.

The occurrence of abnormal behaviors is rare, so the sensing data from IoT devices will form an imbalanced training dataset. For this reason, we
implement the anomaly detection model by first building a normal behavioral representation and then determining abnormal behaviors on the basis of its deviation from the normal representation. In this paper, we use an improved generative adversarial network (GAN), namely WGAN-GP \cite{38-Gulrajani2017}, as the basic anomaly detection model for its powerful ability to capture distribution from high-dimensional complex data.

Similar to GAN, WGAN-GP consists of two models: generator $G$ and discriminator $D$. The basic principle of WGAN-GP is as follows: $G$ generates fake data $\bar {\bm{X}}$ using a random noise $\bm{z}$ that follows a known distribution $P_{\bm{z}}$, and $D$ attempts to distinguish $\bar {\bm{X}}$ from real data $\bm{X}$. $G$ and $D$ are trained through adversarial learning until $G$ can generate fake data $\bar {\bm{X}}$ sharing the same distribution with real data $\bm{X}$, and $D$ can not distinguish between them \cite{39-9863661}. Let $P_{\bm{X}}$ and  $P_{\bar {\bm{X}}}$ represent the distributions of real and fake data, and $D(\bm{X})$ represent the probability that $\bm{X}$ follows the data distribution $P_{\bm{X}}$. The training objective of WGAN-GP is defined as follows:
\begin{equation}
\begin{gathered}
  \mathop {\min }\limits_G \;\mathop {\max }\limits_D \;V(G,D) = \mathop \mathbb{E}\limits_{{\bm{X}} \sim P_{\bm{X}} } [D({\bm{X}})] - \mathop \mathbb{E}\limits_{\bar {\bm{X}} \sim P_{\bar {\bm{X}}} } [D({\bar {\bm{X}}})] \hfill \\
  \;\;\;\;\;\;\;\;\;\;\;\;\;\;\;\;\; + \eta \mathop \mathbb{E}\limits_{\widehat {\bm{X}} \sim P_{\widehat {\bm{X}}}} [(||\nabla _{\widehat X} D(\widehat {\bm{X}})||_2  - 1)^2 ] \hfill \\
\end{gathered},
\end{equation}
where $\eta$ is the coefficient of the gradient penalty, and $\widehat {\bm{X}}$ is defined as the weighted sum of ${\bm{X}}$ and $\bar {\bm{X}}$. During the training process of WGAN-GP, $G$ and $D$ are updated alternately.

WGAN-GP is used as the local anomaly detection model in UAVs. First, we define the discriminator and generator parameters of UAV $n$ as ${\bm w}_n$ and ${\bm \theta} _n$, respectively. In each local update, UAV $n$ performs $N_d$ training iterations on discriminator $D_n$ between two generator training rounds. Assuming that $m$ is the batch size, we sample $m$ random noise $\{ {\bm z}_n^i \} _{i = 1}^m$ to generate fake data $\{ {\bar {\bm X}}_n^i \} _{i = 1}^m$, and sample $m$ real data $\{ {\bm X}_n^i \} _{i = 1}^m$ to train the local discriminator jointly. Hence, for UAV $n$, the loss function of $D_n$ is calculated by
\begin{equation}
\begin{gathered}
LD_n ({\bm w}_n ,{\bm \theta} _n ) =   \hfill \\
- \frac{1}
{m}\sum\limits_{i = 1}^m {\left\{ {D({\bm X}_n^i ) - D({\bar {\bm X}}_n^i ) + \eta [||\nabla _{\widehat {\bm X}_n^i } D(\widehat {\bm X}_n^i )||_2  - 1]^2 } \right\}}  \hfill \\
\end{gathered}.
\end{equation}

Accordingly, at the HAPS, the average loss function of the global discriminator is expressed as follows:
\begin{equation}
\begin{gathered}
  LD({\bm w},{\bm \theta} ) = - \frac{1}
{{m|{\mathcal{N}}_s |}}\sum\limits_{n = 1}^{|{\mathcal{N}}_s |} {\sum\limits_{i = 1}^m {\Big \{ D({\bm X}_n^i ) - D({\bar {\bm X}}_n^i ) + }} \hfill \\
   \;\;\;\;\;\;\;\;\;\;\;\;\;\;\;\;\;\;\;\;\;\;\;\;\;\;\;\;\;\eta[||\nabla _{\widehat {\bm X}_n^i } D(\widehat {\bm X}_n^i )||_2  - 1]^2 \Big \}  \hfill \\
\end{gathered},
\end{equation}
where ${\bm w}$ and ${\bm \theta}$ represent the global discriminator and generator parameters, respectively. ${\mathcal{N}}_S  \subset {\mathcal{N}}$ is the selected UAV subset for global aggregation, and $|{\mathcal{N}}_s |$  is the number of the selected UAVs.

Similarly, the loss function of generator $G_n$ is calculated by
\begin{equation}
LG_n ({\bm w}_n ,{\bm \theta} _n) = - \frac{1}
{m}\sum\limits_{i = 1}^m {\left[ {D({\bar {\bm X}}_n^i )} \right]}.
\end{equation}

Accordingly, at the HAPS, the average loss function of the global generator is expressed as follows:
\begin{equation}
LG({\bm w},{\bm \theta}) = - \frac{1}
{{m|{\mathcal{N}}_s |}}\sum\limits_{n = 1}^{|{\mathcal{N}}_s |} {\sum\limits_{i = 1}^m {\left[ {D({\bar {\bm X}}_n^i )} \right]} }.
\end{equation}

The AFL framework aims to search for the optimal model parameters at the HAPS that minimize the global loss accordingly:
\begin{equation}
\begin{gathered}
  {\bm w}^ *   = \mathop {\arg \min }\limits_{\bm w} LD({\bm w},{\bm \theta} ) \hfill \\
  {\bm \theta} ^ *   = \mathop {\arg \min }\limits_{\bm \theta}  LG({\bm w},{\bm \theta} ) \hfill \\
\end{gathered}.
\end{equation}

Although AFL has distinct privacy advantages like FL, current research shows that sensitive information can still be inferred by using shared parameters during the learning process \cite{28-9522027}. To address this issue, we propose a differentially private WGAN-GP model, where differential privacy is achieved in WGAN-GP by adding carefully designed noise to gradients during the learning process.

\textbf{Definition 1}. A randomized function $F$ is considered as $(\epsilon ,\delta)$-differentially private if the following inequality is satisfied for any two databases ${\bm X}$ and ${\bm {X}}'$ differing in a single point and for any output subset $\bm S$ \cite{40-xie2018differentially}:
\begin{equation}
\Pr (F(\bm X) \in {\bm S}) \leqslant e^\epsilon   \cdot \Pr (F({\bm X}' ) \in {\bm S}) + \delta,
\end{equation}
where $F(\bm X)$ and $F({\bm X}')$ are the outputs of the function $F$ for inputs $\bm X$ and ${\bm X}'$, respectively.

Adding noise to the gradients of the Wasserstein distance is more efficient than adding noise to the final parameters directly with respect to preserving privacy \cite{40-xie2018differentially}. The gradients $\Delta {\bm w}_n$ of discriminator parameter ${\bm w}_n$ after adding noise can be expressed as follows:
\begin{equation}
\Delta {\bm w}_n  =  - \frac{1}
{m}\sum\limits_{i = 1}^m {\{ \nabla _{{\bm w}_n } LD_n^i  + N(0,\sigma _n^2 c_g^2 \bm{I})\} },
\end{equation}
where $\sigma _n$ represents the noise scale, and $c_g$ is the bound on the gradients of the Wasserstein distance.

According to \textbf{Lemma 1} in \cite{40-xie2018differentially}, given the sampling rate $p = \frac{m}{M}$ (where $m$ is the batch size and $M$  is the total number of training data used in each discriminator iteration), the number of discriminator iterations $N_d$ between two generator iterations, and privacy violation $\delta$, then for any positive $\epsilon$, the discriminator parameter guarantees $(\epsilon ,\delta )$-differential privacy with respect to all data used in the generator iteration if we choose
\begin{equation}
\sigma _n  = \frac{{2p\sqrt {N_d \log (\frac{1}
{\delta })} }}
{\epsilon }.
\end{equation}

The training processes of the AFL-based differentially private WGAN-GP model are summarized in Algorithm 1.

\begin{algorithm}[t]
\caption{AFL-Based Differentially Private WGAN-GP.}
\begin{algorithmic}[1]
\REQUIRE The number of time slots $T$; initial generator and discriminator parameters  $\bm{\theta}$ and $\bm{w}$; the number of iterations $N_d$ between two generator training rounds; the number of local iterations $N_l$ per time slot; batch size $m$; Adam hyper-parameters $\alpha ,\beta _1 ,\beta _2$; noise scale $\sigma _n$; bound on the gradient of Wasserstein distance $c_g$.
\ENSURE Converged global model parameters $\bm{\theta}$ and $\bm{w}$.
\FOR{$t = 1:T$}
\STATE Select the UAV subset ${\mathcal{N}}_s (t)$ by CA2C algorithm.
\STATE For each UAV $n$ ($n \in \mathcal{N}$), initialize local parameters by ${\bm{\theta}}_n = \bm{\theta}$, ${\bm{w}}_n = \bm{w}$.
\FOR{$t_1  = 1:N_l$}
\FOR{$t_2  = 1:N_d$}
\STATE Sample a batch of noise $\{ {\bm z}_n^i \} _{i = 1}^m \sim P_{\bm z}$ and a batch of real data $\{ {\bm X}_n^i \} _{i = 1}^m \sim P_{\bm X}$.
\STATE For each UAV $n$ ($n \in {\mathcal{N}}_s (t)$), calculate the loss function $LD_n ({\bm{w}}_n,{\bm{\theta}} _n)$ of $D_n$ according to (6).
\STATE Add noise to the gradients $\Delta {\bm{w}}_n$ of discriminator parameter ${\bm{w}}_n$ according to (12).
\STATE Update ${\bm{w}}_n$ using Adam optimizer:  ${\bm{w}}_n  \leftarrow {\text{Adam(}}\Delta {\bm{w}}_n ,{\bm{w}}_n ,\alpha ,\beta _1 ,\beta _2 )$.
\ENDFOR
\STATE Sample another batch of noise $\{ {\bm z}_n^i \} _{i = 1}^m \sim P_{\bm z}$.
\STATE For each UAV $n$ ($n \in {\mathcal{N}}_s (t)$), calculate the loss function $LG_n ({\bm{w}}_n,{\bm{\theta}} _n)$ of $G_n$ according to (8)
\STATE  Calculate the gradients of generator parameter $\Delta {\bm{\theta}}_n$ by $\Delta {\bm{\theta}} _n  = \frac{1}{m}\sum\limits_{i = 1}^m {\{ \nabla _{{\bm{\theta}} _n } LG_n^i \} }$.
\STATE  Update ${\bm{\theta}}_n$ using Adam optimizer: ${\bm{\theta}}_n  \leftarrow {\text{Adam(}}\Delta {\bm{\theta}}_n ,{\bm{\theta}}_n ,\alpha ,\beta _1 ,\beta _2 )$.
\ENDFOR
\STATE HAPS collects local parameters $\{ {\bm w}_n \} _{n = 1}^{|{\mathcal{N}}_s(t) |}$ and $\{ {\bm \theta}_n \} _{n = 1}^{|{\mathcal{N}}_s(t) |}$ from selected UAVs, and updates the global parameters ${\bm w} = \sum\nolimits_{n \in {\mathcal{N}}_s(t) } {|{\mathcal{X}}_n |{\bm w}_n } /|{\mathcal{X}}|,\;\;{\bm {\theta}}  = \sum\nolimits_{n \in {\mathcal{N}}_s(t) } {|{\mathcal{X}}_n |{\bm {\theta}} _n } /|{\mathcal{X}}|$, where $| {\mathcal{X}}_n |$ is the cardinality of training set ${\mathcal{X}}_n$ and $|{\mathcal{X}}| = \sum\nolimits_{n \in {\mathcal{N}}_s(t) } {|{\mathcal{X}}_n |}$.
\ENDFOR
\end{algorithmic}
\end{algorithm}

\subsection{Communication Model}
The information exchange between UAVs and the HAPS includes the processes of UAVs uploading local models to the HAPS for global aggregation and the HAPS distributing the global model to all UAVs after the averaging operation. We consider line-of-sight (LoS) links for both UAV-HAPS and HAPS-UAV wireless transmission. The LoS path loss in dB between UAV $n$ and the HAPS is modeled by

\begin{equation}
l_n^{{\text{LoS}}} (t) = 20\log (4\pi d_n (t)f_c /c) + h^{{\text{LoS}}},
\end{equation}
where $f_c$ is the carrier frequency, and $d_n (t)$ denotes the distance between UAV $n$ and the HAPS at time $t$. $c$ is the speed of light, and $h^{{\text{LoS}}}$ denotes the mean additional loss of LoS links caused by the free space propagation \cite{1-9453811}.
Hence, the uplink transmission rate from UAV $n$ to the HAPS is given by
\begin{equation}
R_n^u (t) = B^u \log _2 (1 + \frac{{P_n^u 10^{ - l_n^{{\text{LoS}}} (t)/10} }}
{{B^u N_0 }}),
\end{equation}
where $B^u$ is the bandwidth for each uplink channel, which is assumed to be the same for each UAV in this paper. $P_n^u$ is the transmission power of UAV $n$, and $N_0$ denotes the Gaussian noise power spectrum density.

Similarly, the downlink transmission rate from the HAPS to UAV $n$ is given by
\begin{equation}
R_n^d (t) = B^d \log _2 (1 + \frac{{P_n^d 10^{ - l_n^{{\text{LoS}}} (t)/10} }}
{{B^d N_0 }}),
\end{equation}
where $B^d$ is the bandwidth for each downlink channel, which is also assumed to be the same for each UAV. $P_n^d$ is the transmission power that HAPS allocates to UAV $n$.

\subsection{AFL Model Update Latency}
Due to the different quality of local models and different battery power remaining in different UAVs, we try to select a subset of UAVs ${\mathcal N}_s (t) \subset {\mathcal N}$ in each training round to participate in the global aggregation with the goal of minimizing the federated execution time and learning accuracy loss. We introduce $\gamma _n (t)$ to denote whether or not UAV $n$ is selected for global aggregation at time slot $t$, with $\gamma _n (t) = 1$ indicating that UAV $n$ is selected, and $\gamma _n (t) = 0$ indicating otherwise.

The one-round federated execution time includes local model update and upload latency, and global model aggregation and distribution latency.

\emph{1) Local model update latency.} We define $\xi _n^a (t)$ to represent the computation capability of UAV $n$ to execute training tasks at time slot $t$, and $\rho_n$ represents the number of CPU cycles needed to train the local model on one unit of data. Therefore, the local model update latency of UAV $n$ at time slot $t$ is given by
\begin{equation}
T_n^{{\text{update}}} (t) = \frac{{|{\mathcal X}_n (t)|\rho _n }}
{{\xi _n^a (t)}},
\end{equation}
where $|{\mathcal X}_n (t)|$ is the cardinality of dataset ${\mathcal X}_n (t)$.

\emph{2) Local model upload latency.} Let $L(|{\bm \theta} _n | + |{\bm w} _n |)$ denote the total number of bits required by UAV $n$ to upload its local parameters to the HAPS. So for UAV $n$, the local model upload latency is expressed as
\begin{equation}
T_n^{{\text{upload}}} (t) = \frac{{L(|{\bm \theta} _n | + |{\bm w}_n |)}}
{{R_n^u (t)}}.
\end{equation}

\emph{3) Global model aggregation latency.}  We define ${\mathcal N}_s (t) = \{ n|\gamma _n (t) = 1,\forall n\}$ to represent the UAV subset that is selected for global aggregation at time $t$. Assuming that the global model aggregation latency is linearly related to the number of participating UAVs, which is given by
\begin{equation}
T_n^{{\text{aggregation}}} (t) = \varsigma |{\mathcal N}_s (t)|,
\end{equation}
where $|{\mathcal N}_s (t)|$ represents the number of selected UAVs, and $\varsigma$ denotes the unit time for global aggregation.

\emph{4) Global model distribution latency.}   Let $L(|{\bm \theta} | + |{\bm w}|)$ denote the total number of bits required by the HAPS to distribute the global model to UAVs. So the global model distribution latency for UAV $n$ is expressed as
\begin{equation}
T_n^{{\text{distribution}}} (t) = \frac{{L(|{\bm \theta} | + |{\bm w}|)}}
{{R_n^d (t)}}.
\end{equation}

The time cost for AFL is expressed as the maximum one-round execution time in ${\mathcal N}_s (t)$, which is given by
\begin{equation}
\begin{gathered}
  T(t) = \mathop {\max }\limits_{n \in {\mathcal N}_s (t)} (T_n^{{\text{update}}} (t) + T_n^{{\text{upload}}} (t) \hfill \\
  \;\;\;\;\;\;\;\;\;\;\;\;\;\;\;\;\;\;\;\; + T^{{\text{aggregation}}} (t) + T_n^{{\text{distribution}}} (t)) \hfill \\
\end{gathered}.
\end{equation}

Hence, we define the time cost function as
\begin{equation}
\begin{gathered}
  C^T (t) = \frac{1}
{{|{\mathcal N}_s (t)|}}\sum\limits_{n = 1}^{|{\mathcal N}_s (t)|} {(T_n^{{\text{update}}} (t) + T_n^{{\text{upload}}} (t)}  \hfill \\
  \;\;\;\;\;\;\;\;\;\;\;\;\;\;\;\;\;\;\;\;\;\;\;\;\;\; + T^{{\text{aggregation}}} (t) + T_n^{{\text{distribution}}} (t)) \hfill \\
\end{gathered}.
\end{equation}

\subsection{UAV Energy Model}
UAV energy consumption mainly consists of three important parts: propulsion energy, computational energy, and transmission energy.

\emph{1) Propulsion energy.} According to \cite{13-9130055} and \cite{41-8663615}, the propulsion energy of UAV $n$ during a flight is modeled by
\begin{equation}
\begin{gathered}
  E_n^P (t) =  \hfill \\
  \underbrace {\frac{{||{\bm x}_n (t) - {\bm x}_n (t - 1)||}}
{{V_n }}}_{{\text{flying}}\;{\text{time}}} \times \underbrace {\left[ {\kappa _1 V_n^3  + \frac{{\kappa _2 }}
{{V_n }} + \kappa _3 (1 + \frac{{V_n^2 }}
{{g^2 }})} \right]}_{{\text{propulsion power}}} \hfill \\
\end{gathered},
\end{equation}
where $\kappa _1 ,\kappa _2 ,\kappa _3$, and $g$ are constants related to the type of UAVs, and $V_n$ denotes the flying velocity of UAV $n$.

\emph{2) Computational energy.} We define $P_n^c$ as the computational power of UAV $n$. Accordingly, the computational energy consumption of UAV $n$ at time slot $t$ is calculated by
\begin{equation}
E_n^C (t) = \gamma _n (t)P_n^c T_n^{{\text{update}}} (t).
\end{equation}

\emph{3) Transmission energy.} As $P_n^u$ is the transmission power of UAV $n$, the transmission energy consumption of UAV $n$ at time slot $t$ is given by
\begin{equation}
E_n^M (t) = \gamma _n (t)P_n^u T_n^{{\text{upload}}} (t).
\end{equation}

The onboard energy storage of each UAV is limited and represented by $E_n^{{\text{max}}}$. To support the system for $T$ time slots, the total energy consumption of UAV $n$ should satisfy
\begin{equation}
\sum\limits_{t = 1}^T {E_n^P (t) + E_n^C (t) + E_n^M (t) \leqslant } E_n^{{\text{max}}}.
\end{equation}

\subsection{Problem Formulation}
In the three-layer VHetNet, we use UAVs to sense data from IoT devices and employ employ an AFL framework to establish a global anomaly detection model without local data centralization at the HAPS, which can enhance the data privacy and communication efficiency. For this AFL framework, a UAV selection strategy is desirable to select UAVs with high local model quality and low energy consumption to participate in the global aggregation at HAPS. In addition, it is necessary to schedule the trajectory of UAVs and device association to provide the best coverage for IoT devices with minimum energy comsumption. Overall, our objective is to maximize the coverage capacity of UAVs and minimize AFL model execution time and learning accuracy loss over all time slots, which leads to an efficient anomaly detection model. Therefore, we formulate the optimization problem as follows:
\begin{equation}
\begin{gathered}
  \mathop {\min }\limits_{\begin{subarray}{l}
  \{ {\bm x}_n (t)\} ,\{ \lambda _{n,k} (t)\}  \\
  \{ \gamma _n (t)\} ,\{ {\bm w}(t)\}
\end{subarray}}  \frac{1}
{T}\sum\limits_{t = 1}^T {\Big ( - \mu _1 \sum\limits_{n = 1}^N {C_n^C (t)}  + \mu _2 C^T (t)}  \hfill \\
  \;\;\;\;\;\;\;\;\;\;\;\;\;\;\;\;\;\;\;\;\;\;\;\;\;\; + \mu _3 LD({\bm w}(t),{\bm \theta} (t))\Big ) \hfill \\
  s.t.{\text{  a) }}{\bm \theta} (t) = \mathop {\arg \min }\limits_{{\bm \theta} (t)} LG({\bm w}(t),{\bm \theta} (t)) \hfill \\
    \;\;\;\;\;{\text{      b)   }}\lambda _{n,k} (t) \in \{ 0,1\} ,{\text{ }}\gamma _n (t) \in \{ 0,1\} ,{\text{ }}\forall n,k,t \hfill \\
   \;\;\;\;\; {\text{      c)  }}\sum\limits_{n = 1}^N {\lambda _{n,k} (t)}  \leqslant 1,{\text{ }}\forall k,t \hfill \\
    \;\;\;\;\;{\text{      d) }}\sum\limits_{t = 1}^T {E_n^P (t) + E_n^C (t) + E_n^M (t) \leqslant } E_n^{\max } ,\forall n \hfill \\
\end{gathered},
\end{equation}
where $\mu _1$, $\mu _2$, and $\mu _3$ are constant weight parameters used to balance the coverage capacity $\sum\nolimits_{n = 1}^N {C_n^C (t)}$, the execution time $C^T (t)$, and the learning accuracy loss of discriminator $LD({\bm w}(t),\bm{\theta} (t))$. In this system, the learning accuracy losses $LD({\bm w}(t),\bm{\theta} (t))$ and $LG({\bm w}(t),\bm{\theta} (t))$ are measured at the end of each time slot. Since we aim to maximize the coverage capacity, a negative sign is added to the weight parameter $\mu _1$. ${\bm x}_n (t)$ represents the continuous position variables of UAV $n$ at time slot $t$, $\lambda _{n,k} (t)$ represents the binary device association indicator, and $\gamma _n (t)$ represents the binary UAV selection indicator. Constraint (c) indicates that each IoT device can only associate with one UAV to perform sensing tasks. Constraint (d) is used to ensure the total maximum energy consumption.

The optimization problem formulated in (27) is challenging to solve because it is a non-convex combination and NP-hard problem. In addition, due to the time-varying feature of IoT device locations and UAV battery power, it is difficult for traditional optimization algorithms to address this problem. Model-free reinforcement learning (RL) is a promising dynamic programming technique that is capable of handling sequential decision-making processes in dynamic environments \cite{42-franccois2018introduction}. Therefore, we introduce RL to implement the self-scheduling anomaly detection model in the VHetNet-enabled AFL framework.

\section{CA2C-based self-scheduling solution}

\subsection{Modeling of RL Environment}
We model the sequential decision-making problem (27) as a Markov decision process (MDP) represented by $<\mathcal{S},\mathcal{A},\mathcal{P} ,r>$, where $\mathcal{S}$, $\mathcal{A}$, $\mathcal{P}$, and $r$ denote state space, action space, state transition function, and reward, respectively. In the three-layer VHetNet, the HAPS is responsible for observing the dynamic environment and tries to maximize the expected cumulative reward. RL is used to tackle the joint UAV selection, device association, and UAV trajectory planning problem for assisting the implementation of the self-scheduling anomaly detection. According to the system model, the specific elements in the MDP are defined as follows.

\emph{State:} At each time slot $t$, the network state $ {\bm s}(t) \in \mathcal{S}$ consists of the current locations of IoT devices $\{\bm {x}_k (t)\} _{k \in \mathcal{K}}$, UAV locations $\{ {\bm x}_n (t - 1)\} _{n \in \mathcal{N}}$ at last time slot $t-1$, and remaining energy of UAVs $\{ E_n^{\text{Re} } (t)\} _{n \in \mathcal{N}}$. Therefore, the network state at time slot $t$ is expressed as
\begin{equation}
{\bm s}(t) = \{ \{ {\bm x}_k (t)\} _{k \in \mathcal{K}} ,\{ {\bm x}_n (t - 1)\} _{n \in \mathcal{N}} ,\{ E_n^{\text{Re} } (t)\} _{n \in \mathcal{N}} \}.
\end{equation}

\emph{Action:} At time slot $t$, the HAPS is responsible for selecting the corresponding action ${\bm a}(t) \in \mathcal{A}$ based on the observed state $ {\bm s}(t)$, where ${\bm a}(t)$ consists of the UAV locations $\{ {\bm x}_n (t)\} _{n \in \mathcal{N}}$, device association indicators  $\{ \lambda _{n,k} (t)\} _{k \in \mathcal{K} ,n \in \mathcal{N}}$, and UAV selection indicators $\{ \gamma _n (t)\} _{n \in \mathcal{N}}$. The action is expressed as
\begin{equation}
{\bm a}(t) = \{ { \bm x}_n (t)\} _{n \in \mathcal{N}} ,\{ \lambda _{n,k} (t)\} _{k \in \mathcal{K} ,n \in \mathcal{N}} ,\{ \gamma _n (t)\} _{n \in \mathcal{N}} \}.
\end{equation}

\emph{State transition function:} Let $\mathcal{P} ({\bm s}(t + 1)|{\bm s}(t),{\bm a}(t))$ represent the transition probability of the network environment from the current state ${\bm s}(t)$ to a new state ${\bm s}(t + 1)$ after taking an action ${\bm a}(t)$.

\emph{Reward:} Aiming to maximize the coverage capacity of UAVs while minimizing the federated execution time and learning accuracy loss under energy constraints, the instant reward function $r(t)$ is required to evaluate the quality of policy $\bm \pi$ under the current state-action pair $({\bm s}(t),{\bm a}(t))$ \cite{1-9453811}. To obtain the optimal policy ${\bm \pi} ^ *$ under energy constraints, we adopt the ReLu function $f(k) = \max (k,0)$ to calculate the variable $\Theta _n (t)$ with respect to the constraint 27(d) as
\begin{equation}
\Theta _n (t) = \max ((E_n^P (t) + E_n^C (t) + E_n^M (t)) - \frac{{E_n^{\max } }}
{T},0),\;\forall n.
\end{equation}

Therefore, the instant reward function is defined as
\begin{equation}
r(t) =  - (C(t) + \eta ||{\bm \Theta} (t)||_1 ),
\end{equation}
where $C(t) =  - \mu _1 \sum\limits_{n = 1}^N {C_n^C (t)}  + \mu _2 C^T (t) + \mu _3 LD({\bm w}(t),{\bm \theta} (t))$ is the total weighted cost, and $|{\bm \Theta} (t)|$ represents the penalty function. $|{\bm \Theta} (t)|$ is the corresponding vector of $\Theta _n (t)$ and $\|| \cdot ||_1$ is the L1-norm operator.

The objective for the MDP model is to find a policy $\bm \pi$ mapping the state ${\bm s}(t)$ to action ${\bm a}(t)$, i.e., ${\bm a}(t) = {\bm \pi} ({\bm s}(t))$, which is capable of maximizing the total accumulative reward denoted by
\begin{equation}
R(t) = \sum\limits_{t' = t}^T {\chi ^{t' - t} } r(t'),
\end{equation}
where $\chi  \in (0,1)$ is a discount factor indicating how future rewards are important to the current reward.

In the three-layer VHetNet, both state and action spaces are large, so the DRL algorithms that combine deep neural networks (DNNs) and RL are more effective for solving the large-scale decision-making problems. However, the existing DRL algorithms cannot be applied directly to the combinatorial optimization problem formulated in (27). This is because the actions involve both the continuous UAV position variables $\{ {\bm x}_n (t)\} _{n \in \mathcal{N}}$ and the discrete device association $\{ \lambda _{n,k} (t)\} _{k \in \mathcal{K} ,n \in \mathcal{N}}$ as well as the UAV selection indicators $\{ \gamma _n (t)\} _{n \in \mathcal{N}}$, whereas existing DRL algorithms are suitable for problems with purely continuous or purely discrete action spaces.

\subsection{CA2C Algorithm }
To overcome this challenge, we introduce the CA2C algorithm proposed by \cite{32-9154432}, which combines the advantages of the deep deterministic policy gradient (DDPG) method for dealing with continuous decision variables and the deep Q-network (DQN) method for dealing with discrete decision variables.

\emph{DQN:} Traditional Q-learning adopts a Q-table to store all state-action pairs. However, large-scale networks will lead to a large Q-table, and storing all possible state-action pairs in one table is impractical, so DQN is proposed. In DQN, DNNs are utilized to approximate the Q-function, which is represented by $Q({\bm s}(t),{\bm a}(t);{\bm \vartheta} )$, where $\bm \vartheta$ denotes DNN parameters. We train the deep Q-function to achieve the best fitting by minimizing the loss function $L({\bm \vartheta})$ in each iteration, which is defined as the expectation of mean squared error between the estimated Q-value and target value that is given by $L({\bm \vartheta}) = \mathbb{E}[(y(t) - Q({\bm s}(t),{\bm a}(t);{\bm \vartheta} ))^2 ]$. In $L({\bm \vartheta})$, $y(t) = r(t) + \chi \max _{\bm{a}'} Q({\bm s}(t + 1),{\bm a}';{\bm \vartheta} )$ is the target value and ${\bm a}'$ is the action generated by $\varepsilon$-greedy to strike a balance between exploration and exploitation. By $\varepsilon$-greedy, the agent selects a random action ${\bm a}' \in \mathcal{A}$ with probability $\varepsilon$ and selects the best action that follows the greedy policy ${\bm a}' = \arg  \max _{{\bm a}'} Q({\bm s}(t + 1),{\bm a}';{\bm \vartheta} )$ with probability $1-\varepsilon$. Despite the accurate approximation, DNNs may cause a divergence in the training process or ineffective learning due to the non-stationary target values and the correlation among samples. To overcome these difficulties, a pair of techniques are introduced in DQN, namely fixed target network and experience replay \cite{43-mnih2015human}. Hence, the loss function is re-written as
\begin{equation}
\begin{gathered}
L({\bm \vartheta} ) = \mathbb{E}_U [(r(t) + \chi \mathop {\max }\limits_{{\bm a}'} Q(s(t + 1),{\bm a}';{\bm \vartheta} ^ -  )\hfill \\
  \;\;\;\;\;\;\;\;\;\;\;\;\;\;\;\;\;\;\;\;\;\;\;\;\;\;\;\;\;\;\;\;- Q({\bm s}(t),{\bm a}(t);{\bm \vartheta} ))^2 ] \hfill \\
\end{gathered},
\end{equation}
where ${\bm \vartheta} ^ -$ represents the parameters of the target network. At the beginning of the training process, ${\bm \vartheta}$ and ${\bm \vartheta} ^ -$ are initialized with the same values. However, ${\bm \vartheta} ^ -$ only updates with ${\bm \vartheta}$ every $N_u$ steps and keeps unchanged between two individual ${\bm \vartheta}$ updates (slower than the updates of ${\bm \vartheta}$) to avoid a divergence in the training process. $U$ represents the replay memory, and random mini-batches are uniformly sampled from $U$ when performing updates to break the correlation among samples. DQN is only suitable for environments with discrete action spaces. Since the formulated problem in (27) includes both continuous actions (UAV trajectory planning) and discrete actions (device association and UAV selection indicators), it is difficult to solve the problem optimally with the DQN method.

\emph{DDPG:} Policy gradient-based RL approaches can be used to handle the sequential decision-making problems with continuous action spaces by learning deterministic/stochastic policies. These methods aim to optimize a policy using the gradient of the expected reward. To speed up the convergence of policy gradient-based methods, DDPG has been proposed to combine the policy-based method with the value-based method to estimate the policy gradient more efficiently \cite{44-lillicrap2019continuous}. In DDPG, there are two different neural networks, an actor and a critic, with parameters $\bm \varpi$ and $\bm \vartheta$, respectively. For state ${\bm s}(t)$, the actor obtains the continuous action ${\bm a}(t)$ based on a deterministic policy ${\bm \pi} ({\bm s}(t);{\bm \varpi} )$, and the critic evaluates the quality of actions taken via the Q-function $Q({\bm s}(t),{\bm a}(t);{\bm \vartheta} )$. Both the deterministic policy ${\bm \pi} ({\bm s}(t);{\bm \varpi} )$ and the Q-function $Q({\bm s}(t),{\bm a}(t);{\bm \vartheta} )$ are approximated by DNNs. DDPG also uses the target network and the experience replay to facilitate the training process. Sampling random mini-batches from experience replay $U$, the critic is updated by minimizing its loss function, which is expressed as the expectation of a mean squared error between the estimated value $Q({\bm s}(t),{\bm a}(t);{\bm \vartheta} )$ and the target value $y(t) = r(t) + \chi Q({\bm s}(t + 1),{\bm \pi} ({\bm s}(t+1);{\bm \varpi} ^ -  );{\bm \vartheta} ^ -  )$, given by
\begin{equation}
L({\bm \vartheta}) = \mathbb{E}_U [(y(t) - Q({\bm s}(t),{\bm a}(t);{\bm \vartheta} ))^2 ],
\end{equation}
where ${\bm \varpi} ^ - $ and ${\bm \vartheta} ^ - $ represent the parameters of the target actor and the target critic, respectively. The actor is updated using the policy gradient method as follows \cite{44-lillicrap2019continuous}:
\begin{equation}
\nabla _{\bm \varpi}  J({\bm \varpi} ) = \mathbb{E}_U [\nabla _{\bm \varpi}  \pi ({\bm s}(t);{\bm \varpi} )\nabla _{\bm a} Q(s(t),{\bm a}(t);{\bm \vartheta} )].
\end{equation}

\emph{CA2C:} To solve problem (27), whose actions include both the continuous UAV trajectory planning and the discrete device association and UAV selection indicators, we decompose the optimal policy ${\bm \pi} ^ *$ into two parts, i.e., the policy for finding the optimal UAV positions ${\bm \pi} _c^* ({\bm a}_d (t)|{\bm s}(t);{\bm \varpi} )$, and the policy for selecting the optimal device association and UAV selection indicators  ${\bm a}_d^* (t)$ given the optimal UAV positions. More specifically, the optimal discrete policy of device association and UAV selection ${\bm a}_d^* (t)$, which maximizes the Q-value for the network state ${\bm s}(t)$, is determined by
\begin{equation}
{\bm a}_d^* (t) = \mathop {\arg  \max }\limits_{{\bm a}_d (t)} Q^* ({\bm s}(t),[{\bm a}_d (t),{\bm \pi} _c^* ({\bm a}_d (t)|{\bm s}(t);{\bm \varpi} )];{\bm \vartheta} ),
\end{equation}
where policy ${\bm \pi} _c^* ({\bm a}_d (t)|{\bm s}(t);{\bm \varpi} )$ provides the optimal UAV positions given state ${\bm s}(t)$ and discrete policy ${\bm a}_d^* (t)$. The exact estimations of functions $Q^*$ and ${\bm \pi} _c^*$ are time-consuming due to the high-dimensional state and action spaces. To arrive at these estimations, neural networks are adopted to approximate functions $Q^*$ and ${\bm \pi} _c^*$.  Moreover, the training of neural networks is executed by combining the training methods used in DQN and DDPG.

In CA2C, the loss function of the actor is expressed as $J({\bm \varpi} (t)) = \mathbb{E}_U [(Q({\bm s}(t),[{\bm a}_d (t),{\bm \pi} _c ({\bm a}_d (t)|{\bm s}(t);{\bm \varpi} )];{\bm \vartheta} )]$. The actor is updated by applying the policy gradient method as
\begin{equation}
\begin{gathered}
  \nabla _{\bm \varpi}  J({\bm \varpi} ) = \mathbb{E}_U [\nabla _{\bm \varpi}  {\bm \pi} _c ({\bm a}_d (t)|{\bm s}(t);{\bm \varpi} ) \hfill \\
   \times \nabla _{{\bm a}_{_c } } Q({\bm s}(t),[{\bm a}_d (t),{\bm a}_c (t)];{\bm \vartheta} )|_{{\bm a}_c (t) = \pi _c ({\bm a}_d (t)|{\bm s}(t);{\bm \varpi} )} ] \hfill \\
\end{gathered}.
\end{equation}

\begin{algorithm}[t]
\caption{Training Processes for Critic.}
\begin{algorithmic}[1]
\REQUIRE Sample $I$  experience samples ${\bm e}_i  = (\tilde {\bm s},\tilde {\bm a},\tilde r,\tilde {\bm s}')|_{i = 1}^I$ from replay buffer $U$; critic parameter $\bm \vartheta$; target critic parameter ${\bm \vartheta} ^ -$; soft update parameter $\tau$.
\ENSURE Updated parameters $\bm \vartheta$ and ${\bm \vartheta} ^ -$.
\STATE Obtain the UAV location ${\bm \pi} _c (\tilde {\bm a}_d |\tilde {\bm s}';{\bm \varpi} )$ for state $\tilde {\bm s}'$.
\STATE Determine device association and UAV selection actions ${\bm a}'_d$ for state $\tilde {\bm s}'$ based on the Q-value estimated by the target critic.
\STATE Target actor calculates the UAV location ${\bm a}'_c$ for state $\tilde {\bm s}'$.
\STATE According to (34), calculate $y$ by adding ${\tilde r}$ and the output Q-value $Q(\tilde {\bm s}',[{\bm a}'_d ,{\bm \pi} _c ({\bm a}'_d |\tilde {\bm s}';\bm{\varpi} ^ -  )];\bm{\vartheta} ^ -  )$ of target critic.
\STATE Update $\bm \vartheta$ by using the Adam optimizer in the critic.
\STATE Update ${\bm \vartheta} ^ -$ in the target critic via soft update ${\bm \vartheta} ^ -   = (1 - \tau ){\bm \vartheta} ^ -   + \tau {\bm \vartheta}$.
\end{algorithmic}
\end{algorithm}

\begin{algorithm}[t]
\caption{Training Processes for Actor.}
\begin{algorithmic}[1]
\REQUIRE Sample $I$  experience samples  ${\bm e}_i  = (\tilde {\bm s},\tilde {\bm a},\tilde r,\tilde {\bm s}')|_{i = 1}^I$ from replay buffer $U$; actor parameter $\bm \varpi$; target actor parameter ${\bm \varpi} ^ -$; soft update parameter $\tau$.
\ENSURE Updated parameters $\bm \varpi$ and ${\bm \varpi} ^ -$.
\STATE According to (37), calculate the gradients of Q-value $Q(\tilde {\bm s},[\tilde{\bm a}_d ,\tilde{\bm a}_c];{\bm \vartheta} )$ with respect to the UAV locations for all the sampled experiences.
\STATE Update $\bm \varpi$ by using the Adam optimizer in the actor.
\STATE Update ${\bm \varpi} ^ -$ in the target actor via soft update ${\bm \varpi} ^ -   = (1 - \tau ){\bm \varpi} ^ -   + \tau {\bm \varpi}$.
\end{algorithmic}
\end{algorithm}

\begin{algorithm}[t]
\caption{CA2C-based Self-Scheduling Algorithm.}
\begin{algorithmic}[1]
\REQUIRE Initial parameters $\bm \vartheta$, $\bm \varpi$, ${\bm \vartheta} ^ -$, ${\bm \varpi} ^ -$; replay buffer $U = \emptyset$; exploration parameter $\varepsilon$; the maximum number of training episodes $N_{{\text{ep}}}$; the number of time slots $T$; mini-batch size $I$.
\ENSURE Converged model parameters $\bm \vartheta$ and $\bm \varpi$.
\FOR{$episode = 1:N_{{\text{ep}}}$}
\STATE Receive initial observation state $\bm s$.
\FOR{$t  = 1:T$}
\STATE With probability $\varepsilon$, choose random device association and UAV selection actions ${\bm a}_d (t)$, otherwise choose ${\bm a}_d (t)$ according to (36).
\STATE Determine the UAV locations for the selected device association and UAV selection actions as ${\bm a}_c (t) = {\bm \pi} _c ({\bm a}_d (t)|{\bm s}(t);{\bm \varpi} )$.
\STATE Transition to a new state ${\bm s}(t+1)$ and get instant reward $r(t)$.
\STATE Store experience ${\bm s}(t)$, ${\bm a}(t) = [{\bm a}_d (t),{\bm a}_c (t)]$, $r(t)$ and ${\bm s}(t+1)$ in replay buffer $U$ as ${\bm e}_i  = (\tilde {\bm s},\tilde {\bm a},\tilde r,\tilde {\bm s}')$.
\STATE Sample $I$ experience samples ${\bm e}_i  = (\tilde {\bm s},\tilde {\bm a},\tilde r,\tilde {\bm s}')|_{i = 1}^I$ from replay buffer $U$.
\STATE Train the critic according to Algorithm 1.
\STATE  Train the actor according to Algorithm 2.
\ENDFOR
\ENDFOR
\end{algorithmic}
\end{algorithm}

As for the training process of the critic network, the corresponding loss function can be expressed as follows:

\begin{equation}
\begin{gathered}
  L({\bm \vartheta} ) = \mathbb{E}_U [(y(t) - Q({\bm s}(t),{\bm a}(t);{\bm \vartheta} ))^2 ] \hfill \\
  y(t) = r(t) + Q({\bm s}(t + 1),[{\bm a}'_d ,{\bm \pi} _c ({\bm a}'_d |{\bm s}(t + 1);{\bm \varpi} ^ -  )];{\bm \vartheta} ^ -  ) \hfill \\
  {\bm a}'_d  = \mathop {\arg \max }\limits_{{\bm a}'_d } Q({\bm s}(t + 1),\;[{\bm a}'_d ,{\bm \pi} _c ({\bm a}'_d |{\bm s}(t + 1);{\bm \varpi} )];{\bm \vartheta} ) \hfill \\
\end{gathered}.
\end{equation}

In (38), we choose the action for the next state according to the trained networks (i.e., the actor and the critic, with parameters ${\bm \varpi}$ and ${\bm \vartheta}$, respectively), while we estimate the Q-value for the next state using target networks (i.e., the target actor and the target critic, with parameters ${\bm \varpi} ^ -$  and ${\bm \vartheta} ^ -$, respectively) \cite{32-9154432}. Algorithm 2 and 3 summarize the training processes of the actor and critic, respectively. Also, the detailed steps of the CA2C algorithm are presented in Algorithm 4.

\section{Simulation Results and Analysis}
In this section, we conduct simulations to evaluate the performance of the proposed self-scheduling anomaly detection scheme for ubiquitous IoT. In doing so, the proposed CA2C-AFL approach is implemented in pytorch1.11 (Python 3.7) and carried out by a computer with a CPU capacity of 12 Intel(R) Core(TM) i7-10750H CPU 2.6 GHz and a RAM of 16 GB. We compare the proposed CA2C-AFL approach with the following three approaches:

\emph{1) DQN-AFL:} This approach uses DQN to implement the discrete device association and UAV selection strategy, and the continuous UAV trajectory planning is implemented in a random manner.

\emph{2) DDPG-FL:} This approach adopts DDPG to plan the UAV trajectory, and the device association strategy is implemented on the basis of the minimum distance. Moreover, no UAV selection process is included, so all UAVs will participate in the training of the FL-based anomaly detection model.

\emph{3) Standalone:} None of the strategies for device association, UAV selection, or UAV trajectory planning are included. Each UAV trains its own anomaly detection model without any data or information exchange.

\subsection{Experiment Scenario}
For our simulations, we consider a VHetNet that consists of a HAPS and five airborne UAVs to support a coverage area of 1 km $\times$ 1 km, where 20 target devices are randomly located. The transmission power of the HAPS and each UAV is set as 33 dBm and 26 dBm, respectively. The total energy storage of each UAV is 50 kJ and the computation capability of each UAV is assumed to be 80,000 cycles/s. The bandwidth of the uplink and downlink between the HAPS and each UAV is set as 5 MHZ and 20 MHZ, respectively.

The size of the transmitted data for the model parameters is 5 kbits, and the unit computation power is set as 5 W. The batch size for each local training is assumed to be 256, and the number of CPU cycles needed for training the local model on one batch data is 4 cycles/s. Other relevant simulation parameters are listed in Table I, where the UAV setting parameters follow the UAV energy model established in \cite{41-8663615}.

\begin{table}[!t]
\centering
\caption{Simulation parameters}
\renewcommand\arraystretch{1.32}
\scalebox{0.8}{\begin{tabular}{c c c}
\hline
\hline
Parameters&Description&Value\\
\hline
$f_c$ & Carrier frequency &2 GHz\\
$c$ & Speed of light & $3\times10^{8}$ m/s\\
$h^{{\text{LoS}}}$ & Mean additional loss for LoS links&10 dB \\
$N_0$ & Gaussian noise power spectrum density &-174 dBm/Hz\\
$\xi$ & A parameter reflecting the sensing performance & $1.07\times10^{-4}$\\
$m$ &Batch size for WGAN-GP training &20\\
$N_d$ &\thead{The number of discriminator updates\\ between two generator iterations} &6\\
$N_l$ & The number of local iterations per time slot &20\\
$P_{{\text{th}}}$ &Minimum successful sensing probability&0.9\\
$\kappa _1,\kappa _2,\kappa _3$ &UAV type parameters &0.009, 357, 80\\
$g$ &UAV type parameter&69\\
$V_n$ &UAV flying velocity&30 m/s\\
$\mu _1,\mu _2,\mu _3$ &Constant weight parameters&2, 4, 10\\
$\eta$ &Penalty coefficient for energy& 0.01\\
$\chi$ &Discount factor&0.9\\
$\varepsilon$ &Probability for exploration in greedy rule&0.1\\
$U$ &The capacity of replay buffer&2000\\
$T$ &Number of time slots in each episode&50\\
$N_{{\text{ep}}}$ &Maximum number of training episodes&200\\
\hline
\end{tabular}}
\end{table}

\emph{Dataset}: We choose a well-known dataset published by Inter Berkeley Research Lab \cite{45-28584714} as sensing data from ground IoT devices. The dataset includes temperature, humidity, light, and voltage features collected from 54 distributed Mica2Dot sensors every 31 seconds during the period of February 28th, 2004 to April 5th, 2004. Each UAV dataset is built on the sensing data from its associated IoT devices. We divide each UAV subset into three disjoint datasets: a training dataset for model training, a validation dataset for setting the discriminant criterion of anomalies, and a test dataset to evaluate performance \cite{39-9863661}.

\subsection{Convergence Comparison}
We first evaluate the convergence performance of the proposed CA2C approach with different learning rates, which are set as 0.01, 0.001, 0.0001, and 0.00001, sequentially. Fig. 2 shows the convergence of the system cost during the training processes of the CA2C approach. The system cost is the objective function defined in (27), which is the weighted combination of the coverage capacity, the execution time, and the learning accuracy loss. Note that an episode includes 50 time slots. Learning rates have a major impact on system cost and convergence speed. From Fig. 2, we can see that although a larger learning rate will mean a faster convergence process, it will lead to a higher system cost. However, the smaller learning rate does not always lead to less system cost. As shown in Fig. 2, when the learning rate is smaller than 0.0001 (i.e., when the learning rate is equal to 0.00001), the convergence process becomes slower, but the achieved system cost is not smaller than the situation when the learning rate is set as 0.0001. Therefore, we set the learning rate as 0.0001 to account for both the convergence speed and the system cost.
\begin{figure}[!t]
\centering
\includegraphics[width=2.8in]{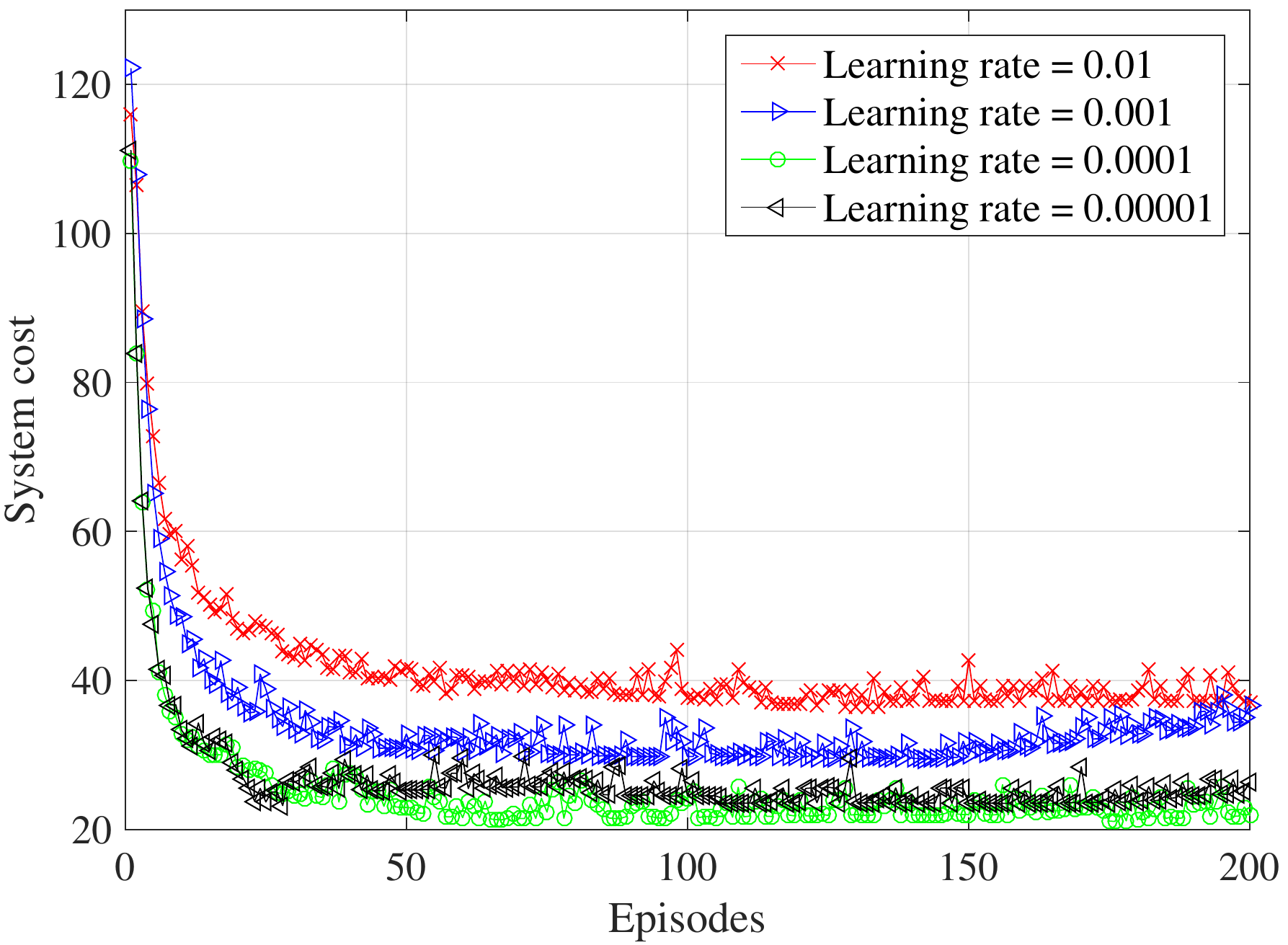}
\caption{Convergence of the CA2C approach in terms of system cost with different learning rates.}
\end{figure}

Fig. 3 compares the convergence of the proposed differentially private WGAN-GP-based anomaly detection model under four different training frameworks, which are built on the proposed CA2C-AFL approach, the DQN-AFL approach, the DDPG-FL approach, and the standalone approach, respectively. From Fig. 3, we can observe that both the discriminator loss and generator loss can converge after certain episodes, which indicates that the generator and discriminator can be successfully trained in all four training frameworks. However, under the standalone training framework, the convergence speed is slow, and the training process is hard to converge to a stable value. This is because the standalone training framework trains the proposed model in each UAV without any data or information exchange, which causes poor data abundance and model robustness. Because the proposed CA2C-AFL-based training framework will plan the best trajectory for UAVs and select the best UAV subset with high model quality in the training process, it achieves the best convergence performance compared to the other three training frameworks.
\begin{figure*}[htbp]
    \centering
    \subfigure[CA2C-AFL]{\label{Fig:R1}
    \includegraphics[width=1.72in]{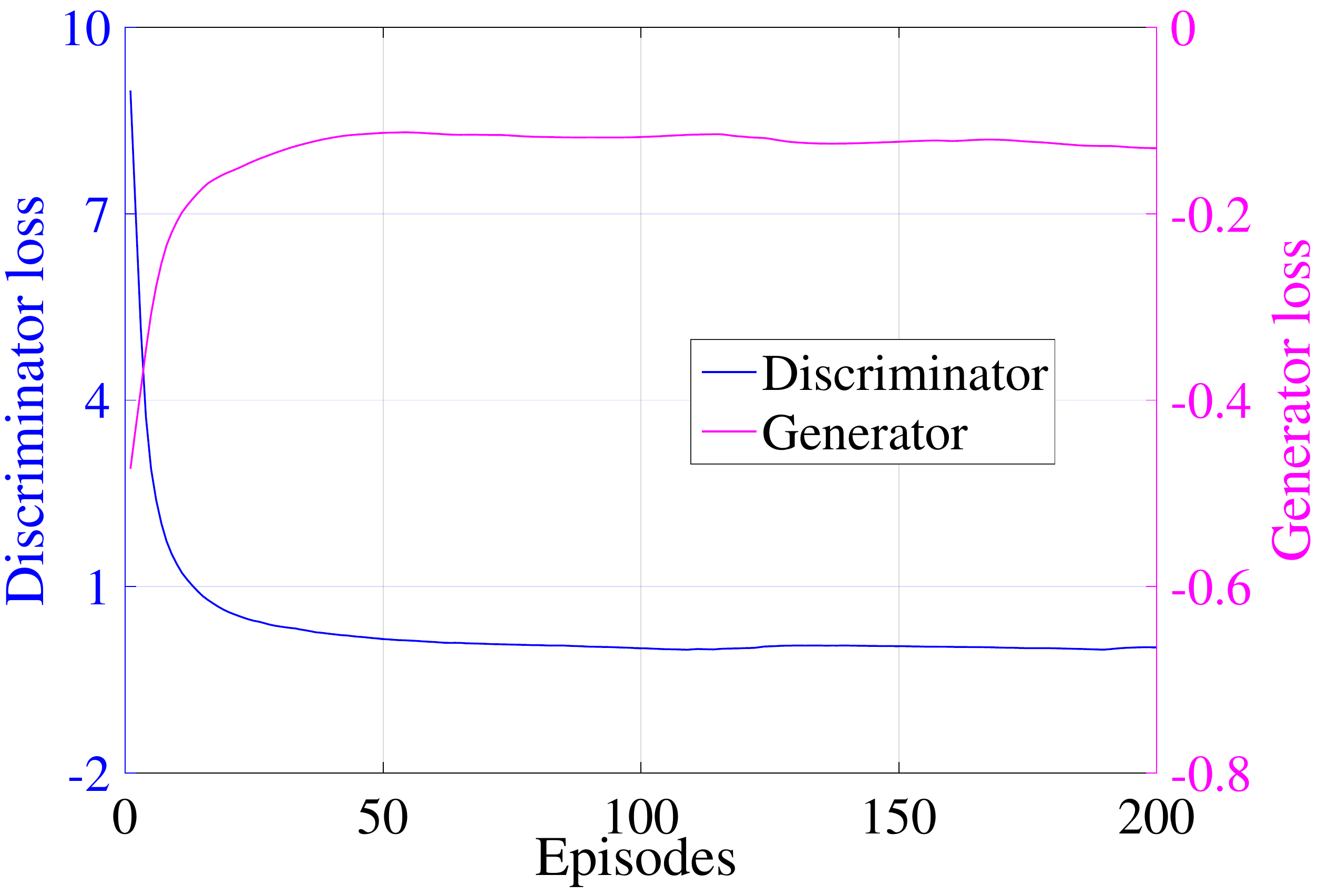}}
    \subfigure[DQN-AFL]{\label{Fig:R2}
    \includegraphics[width=1.72in]{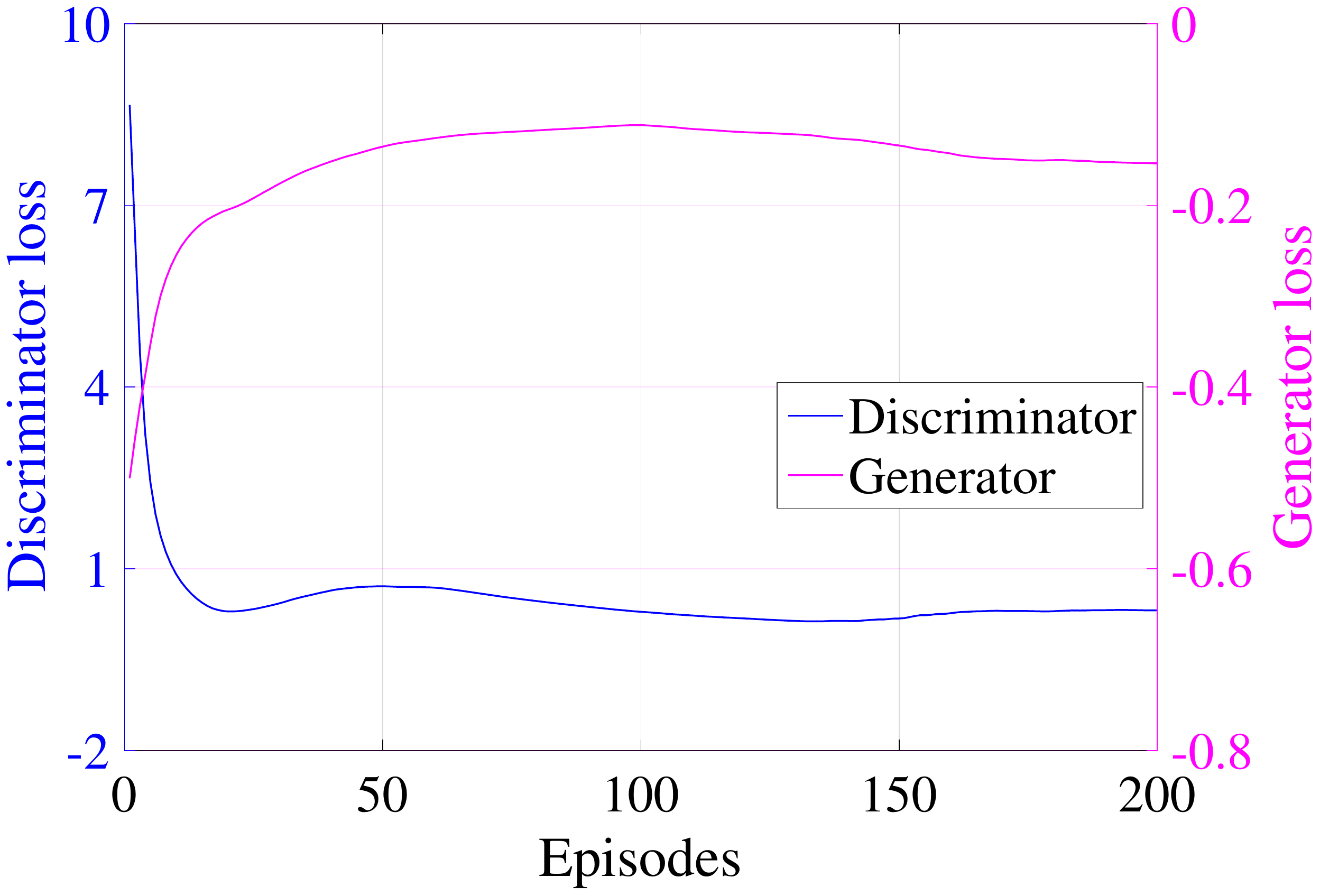}}
    \subfigure[DDPG-FL]{\label{Fig:R3}
    \includegraphics[width=1.72in]{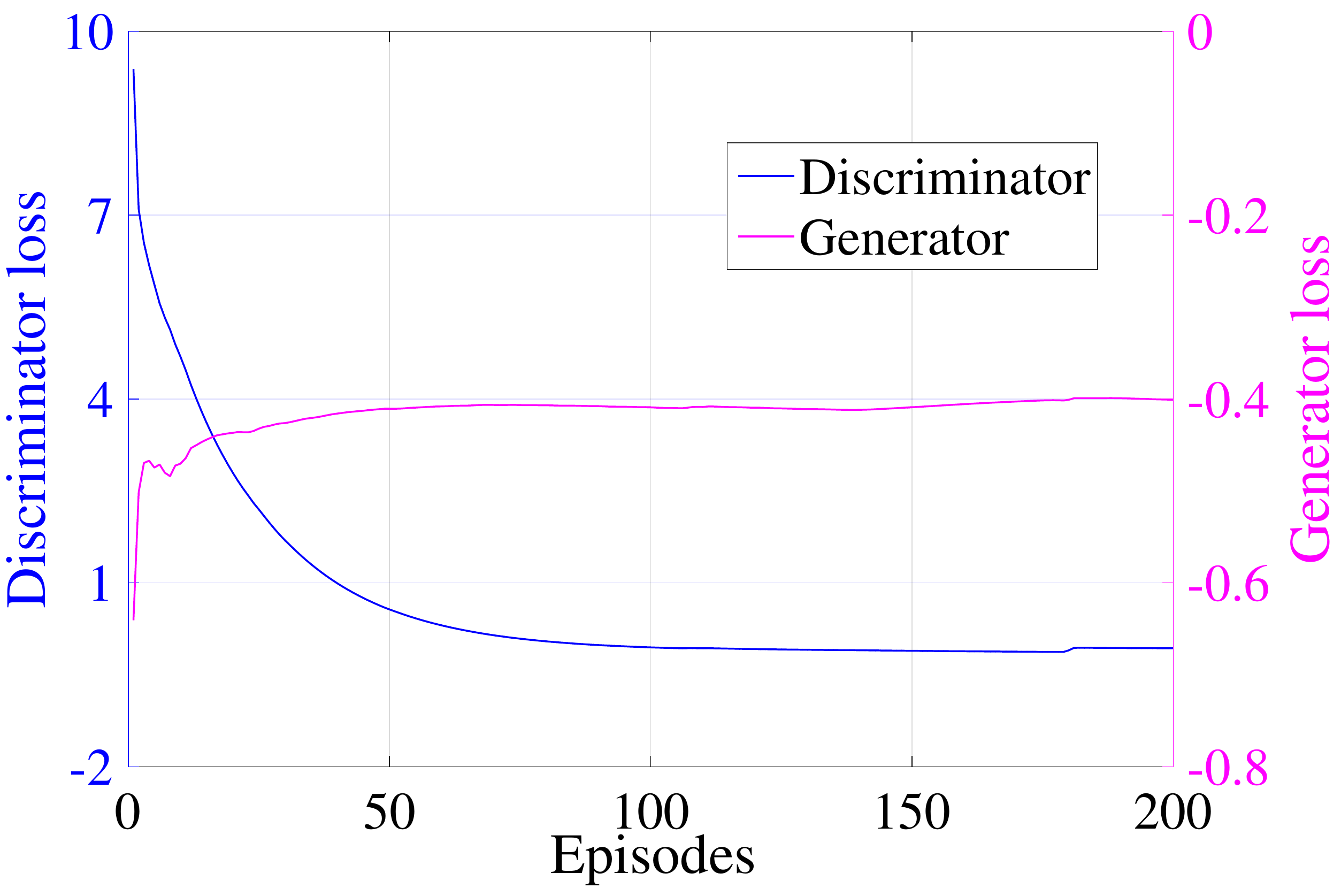}}
    \subfigure[Standalone]{\label{Fig:R4}
    \includegraphics[width=1.72in]{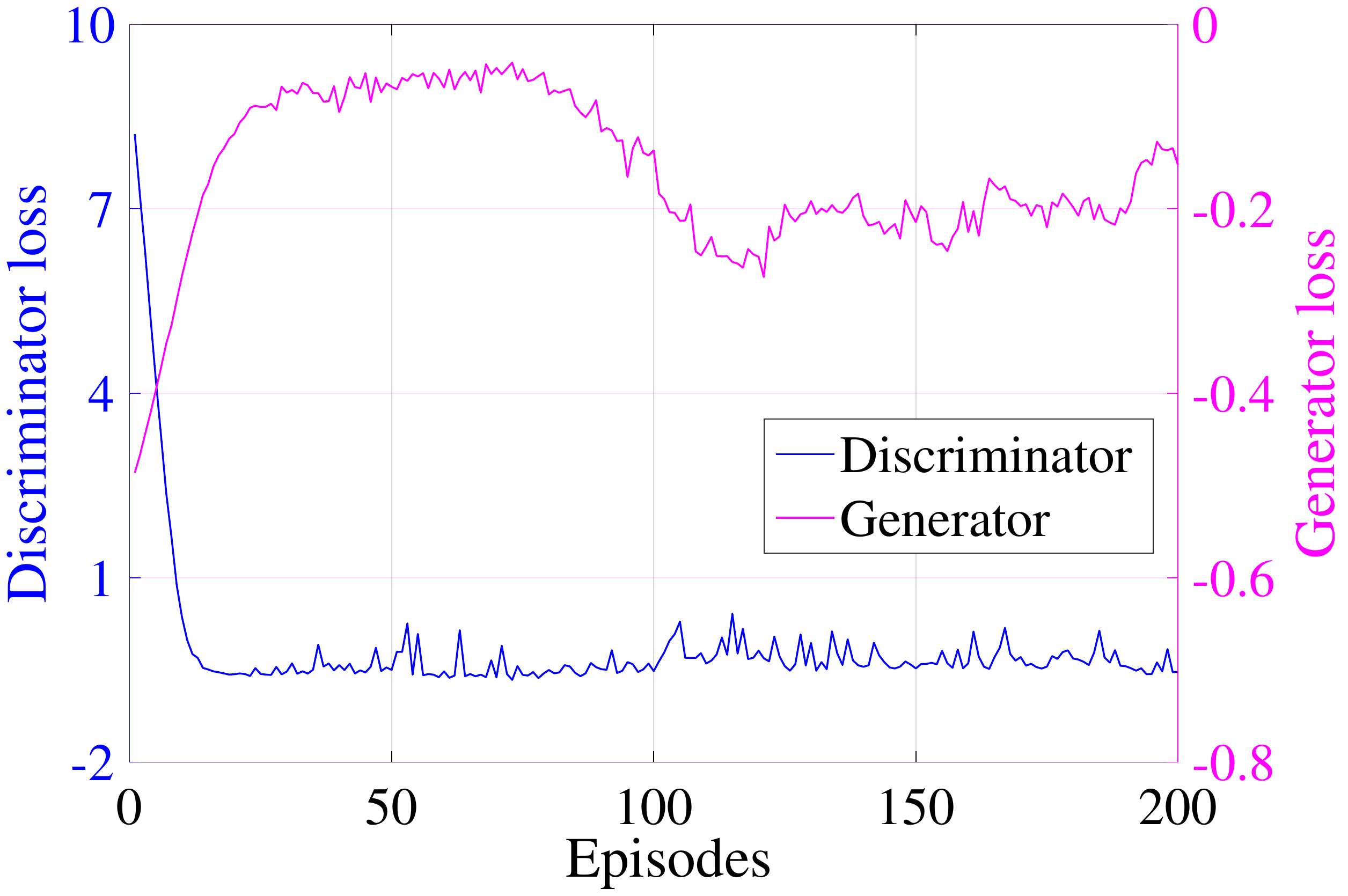}}
    \caption{Convergence of the differentially private WGAN-GP model under different training frameworks.}
\end{figure*}

\subsection{Anomaly Detection Performance}
There are three basic evaluation metrics used to reflect the performance of different anomaly detection algorithms, namely precision, recall, and F1-score, whose meanings are explained as follows:

\emph{Precision:} The ratio of correctly identified abnormal behaviors relative to all behaviors identified as abnormal.

\emph{Recall:} The ratio of correctly identified abnormal behaviors relative to all behaviors set as abnormal in advance.

\emph{F1-score:} The weighted average of precision and recall, given by $F1\textrm{-}score  = (precision \times recall)/(precision + recall{\text{)}}$.

Once the proposed CA2C-AFL-based anomaly detection model is trained well, it can be used to identify abnormal behaviors by calculating their anomaly scores, which are defined as the weighted combination of the discriminator loss and generator loss \cite{39-9863661}. If the anomaly score of a new behavior is greater than a preset $threshold$, it will be determined as abnormal. The relationship between precision, recall, and F1-score with different $threshold$ values is depicted in Fig. 4, where each point represents a different $threshold$ value. As we can see, the $threshold$ values have a major impact on the performance of the proposed anomaly detection model. Therefore, we introduce a simple method to achieve an effective $threshold$ based on the validation dataset. Random noise is injected into the validation dataset to simulate abnormal behaviors. Then, we calculate the average anomaly scores $A_{{\text{normal}}}$ and $A_{{\text{abnormal}}}$ of the normal and abnormal data in the modified validation dataset. Finally, the threshold for the anomaly score is defined as ${{(A_{{\text{normal}}}  + A_{{\text{abnormal}}} )} \mathord{\left/
 {\vphantom {{(A_{{\text{normal}}}  + A_{{\text{abnormal}}} )} 2}} \right.\kern-\nulldelimiterspace} 2}$.
\begin{figure}[!t]
\centering
\includegraphics[width=2.5in]{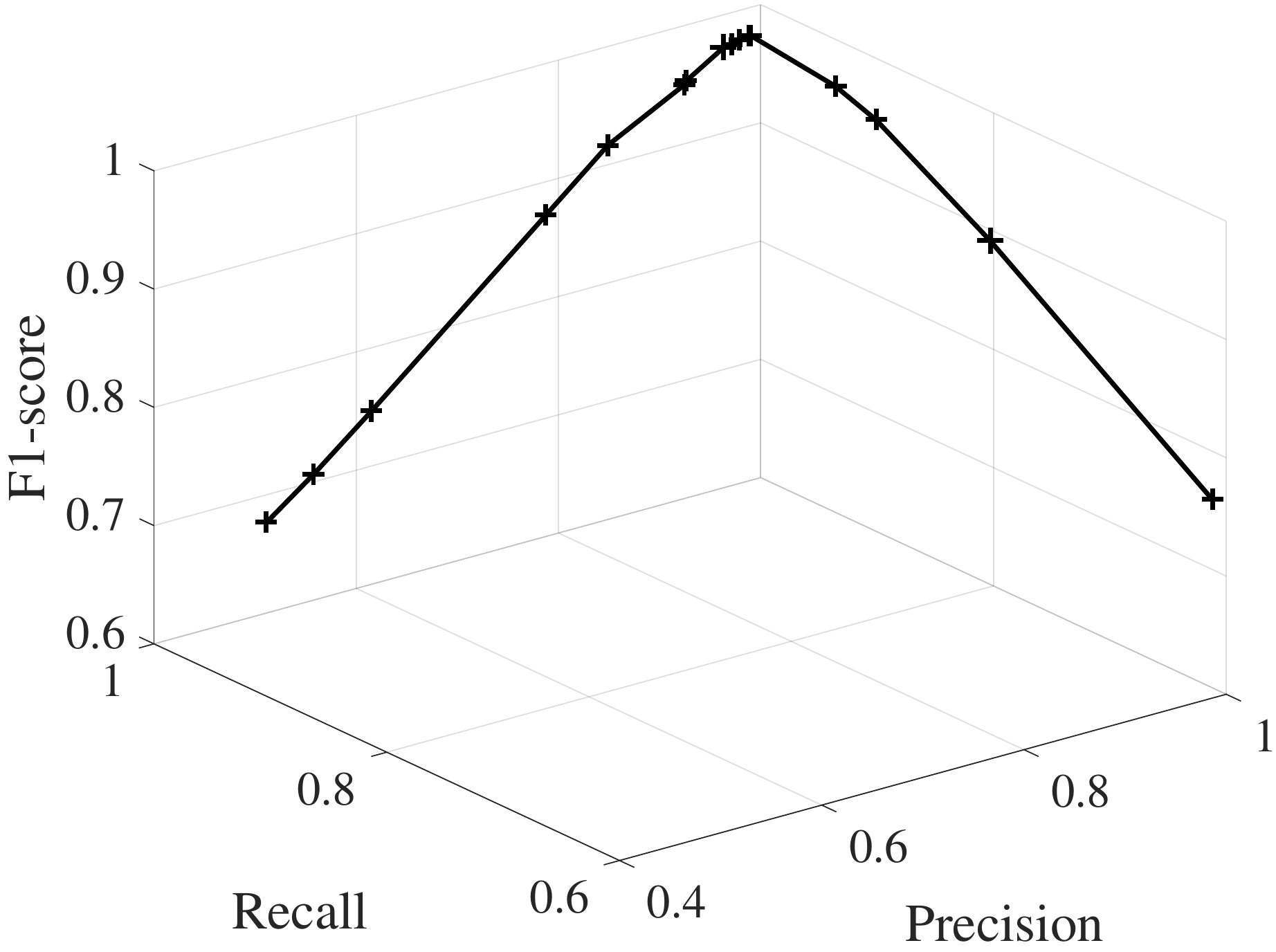}
\caption{The relationship among precision, recall, and F1-score.}
\end{figure}

\begin{figure}[!t]
    \centering
    \subfigure[Precision]{\label{Fig:R1}
    \includegraphics[width=2.1in]{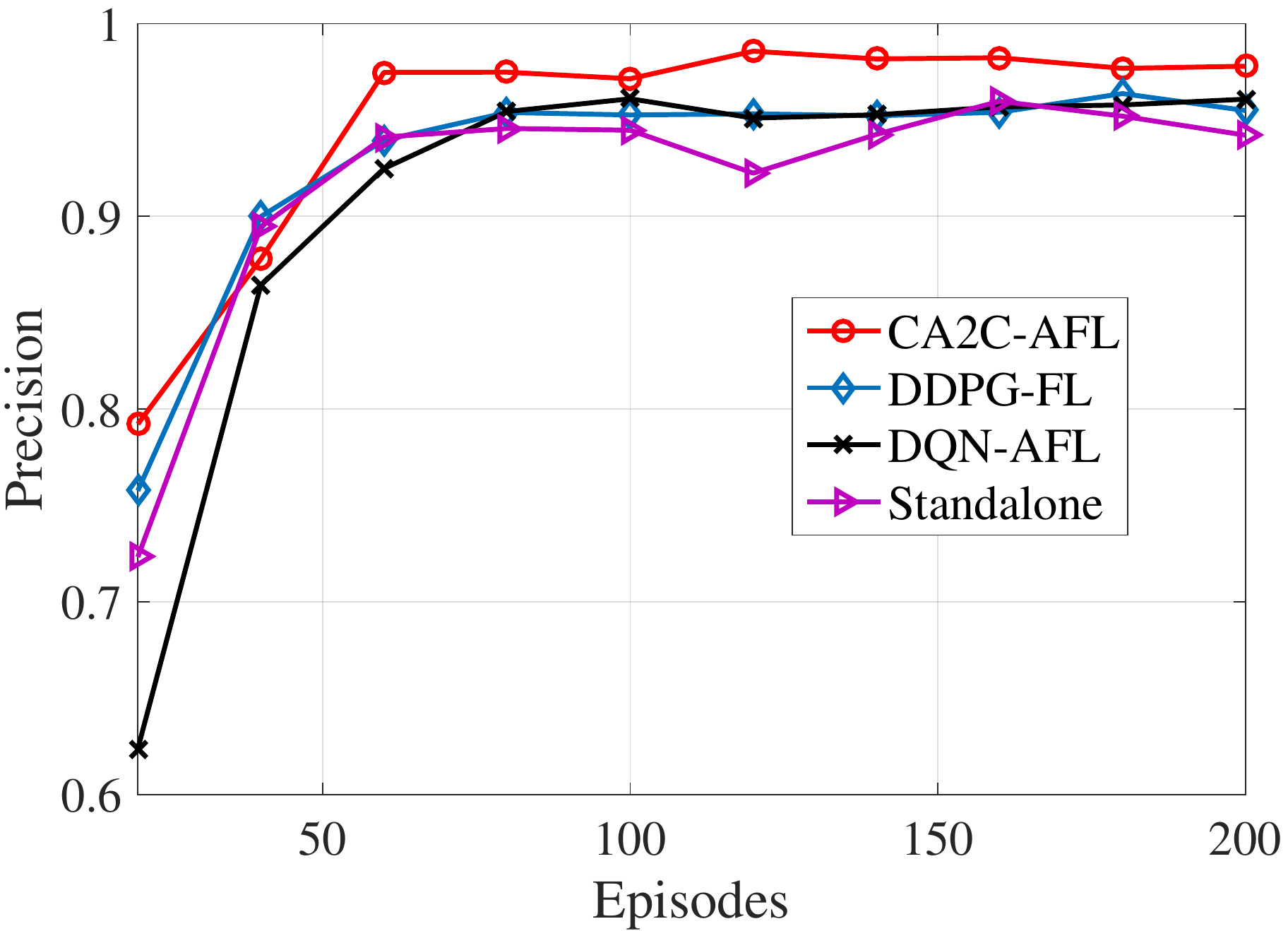}}
    \subfigure[Recall]{\label{Fig:R2}
    \includegraphics[width=2.1in]{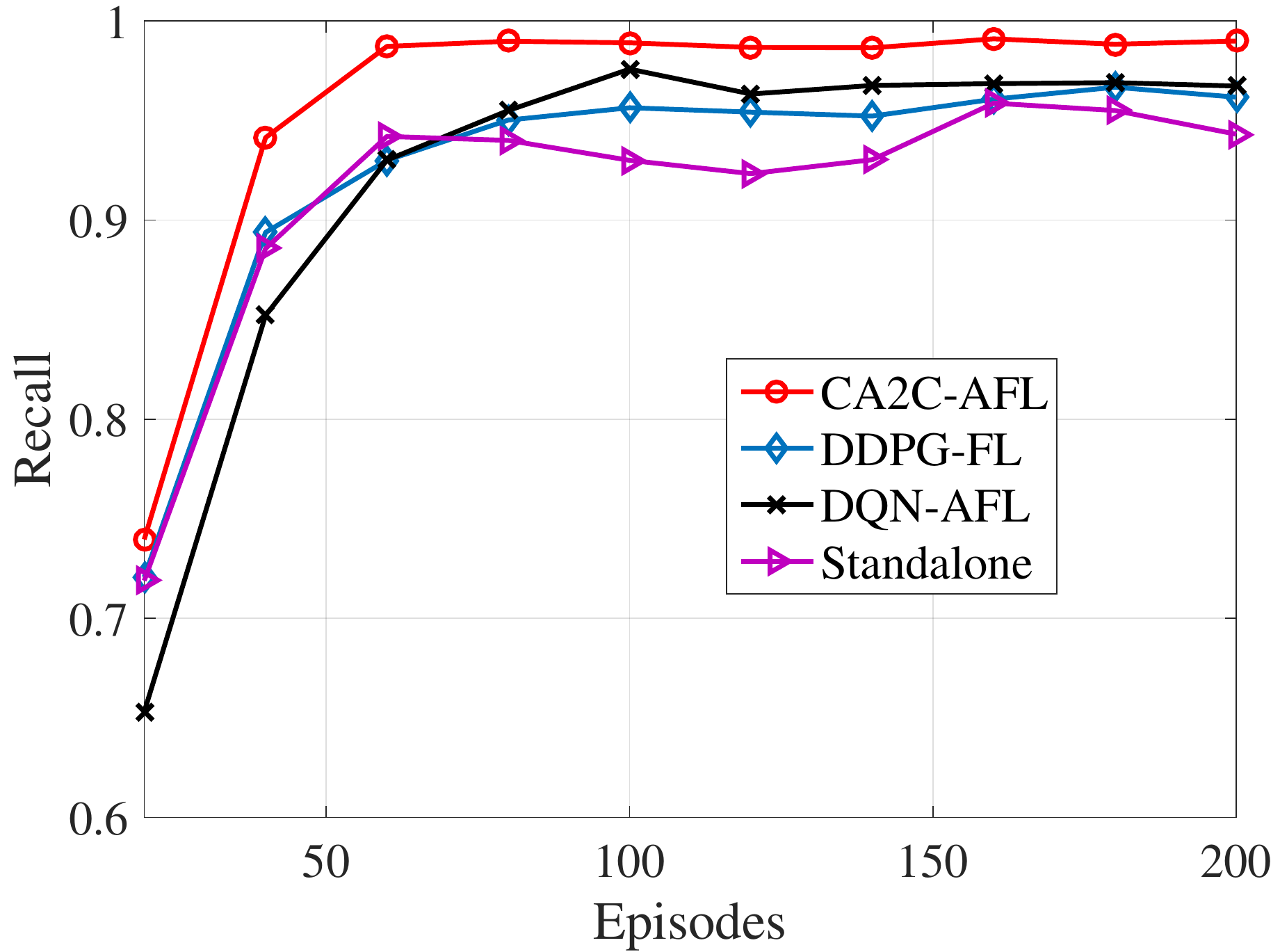}}
    \subfigure[F1-score]{\label{Fig:R3}
    \includegraphics[width=2.1in]{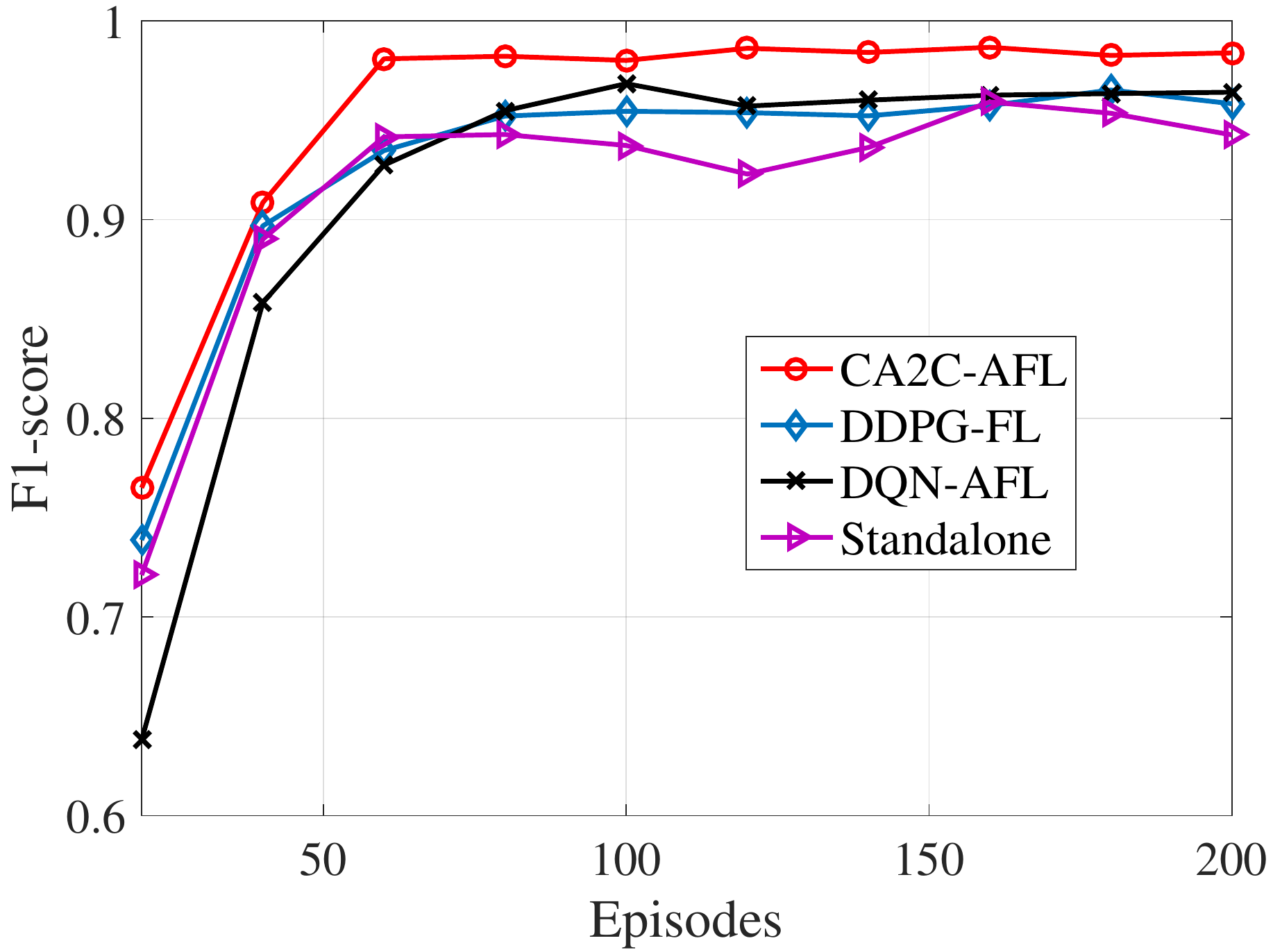}}
    \caption{Detection performance comparison among four different algorithms.}
\end{figure}

Fig. 5 shows the detection performance of four different algorithms: namely the proposed CA2C-AFL-based, the DDPG-FL-based, the DQN-AFL-based, and the standalone anomaly detection algorithms. Precision, recall, and F1-score are used to evaluate the performance of different algorithms. We can observe that the convergence trend of the three evaluation metrics is nearly consistent with that of the discriminator loss and generator loss, which is shown in Fig. 3. It indicates that the four anomaly detection algorithms can capture the accurate data distribution  from the training dataset with enough training episodes. From Fig. 5, we can see that the proposed CA2C-AFL algorithm can achieve the best detection performance in terms of all the three evaluation metrics. This is because the proposed algorithm will plan an optimal trajectory for UAVs and select UAVs with the highest quality models to join the global aggregation in the training process. The detection performance of the DDPG-FL and DQN-AFL algorithms is inferior to our proposed algorithm because these two include either UAV trajectory planning or UAV selection, but not both as our algorithm does. Besides, the detection performance of the standalone algorithm is the most unstable and inaccurate due to the lack of information fusion and exchange.

To further compare the detection performance of the four different algorithms, we use a new dataset, one that is not connected with any training process, as the test dataset to evaluate their generalization ability. From Fig. 6, we can see that the proposed  CA2C-AFL algorithm can achieve the best detection accuracy when facing a new and unknown dataset. The DDPG-FL and DQN-AFL algorithms are inferior to the proposed algorithm but are still in an acceptable range. These results are due to the three algorithms including a process of information fusion and exchange among multiple datasets, which can enhance the generalization ability of the trained models. Comparing the detection performance of the standalone algorithm shown in Fig. 5 and Fig. 6, we can observe that its detection accuracy decreases greatly, which means that the trained model has difficulty distinguishing normal behaviors from abnormal ones when presented with a new dataset. Therefore, the anomaly detection model based on the standalone algorithm has the worst generalization ability due to the lack of data diversity in its training process.
\begin{figure}[!t]
\centering
\includegraphics[width=2.8in]{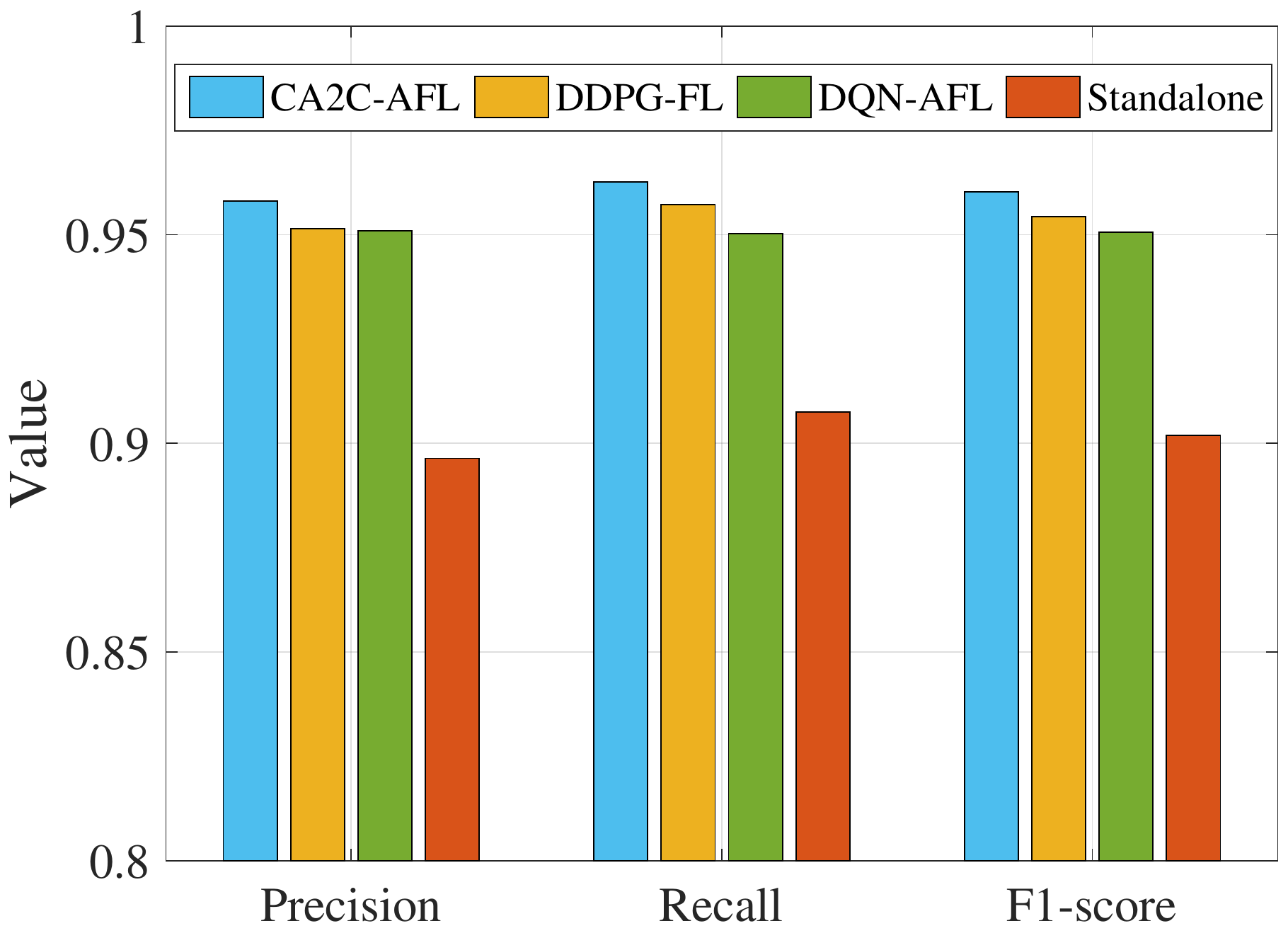}
\caption{ Detection performance comparison among four different algorithms with a new dataset.}
\end{figure}

\subsection{Energy Consumption and Learning Delay}
We also compare the average UAV energy consumption of the proposed CA2C-AFL algorithm with the DDPG-FL, DQN-AFL, and standalone algorithms. Fig. 7 shows the average energy consumption of all UAVs and their mean values. The UAV energy consumption presented in Fig. 7 is averaged over the total 200 episodes. From Fig. 7, we can see that the proposed CA2C-AFL algorithm consumes the least amount of energy among the four algorithms. The reason is that the proposed algorithm will plan the UAV trajectory with the aim of saving energy. Besides, unlike FL, in AFL, just a subset of UAVs need to update their local models in one global training round, which will also decrease the energy consumption. Since the random flying trajectory strategy adopted by the DQN-AFL and standalone algorithms will cause more energy consumption, the DDPG-FL algorithm consumes less energy than either of those algorithms. The standalone algorithm consumes more energy than the DQN-AFL algorithm because it requires all UAVs to update their own models in each training round.
\begin{figure}[!t]
\centering
\includegraphics[width=3.5in]{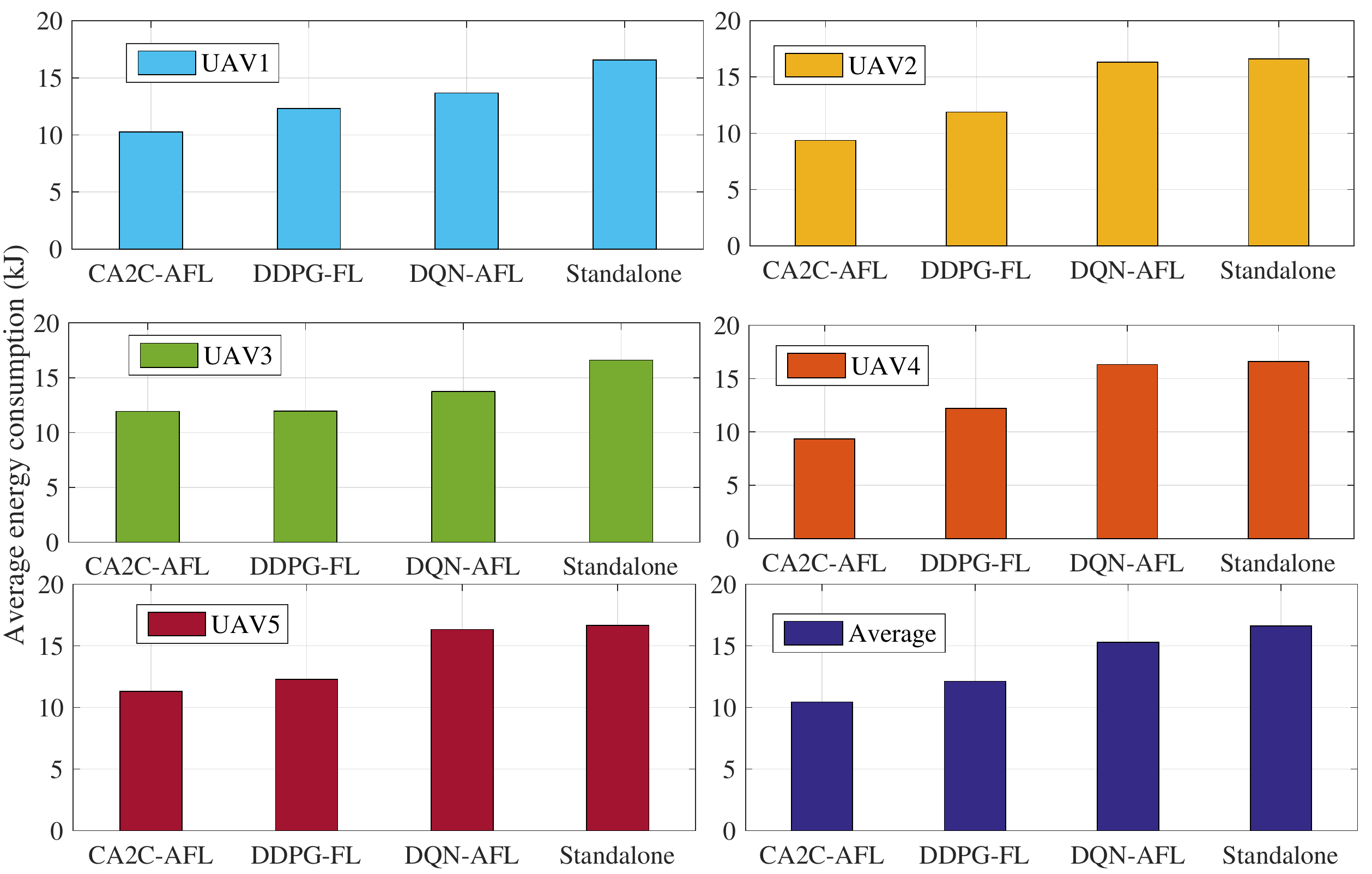}
\caption{Comparison of average UAV energy consumption among four different algorithms.}
\end{figure}

\begin{figure}[t]
\centering
\includegraphics[width=3in]{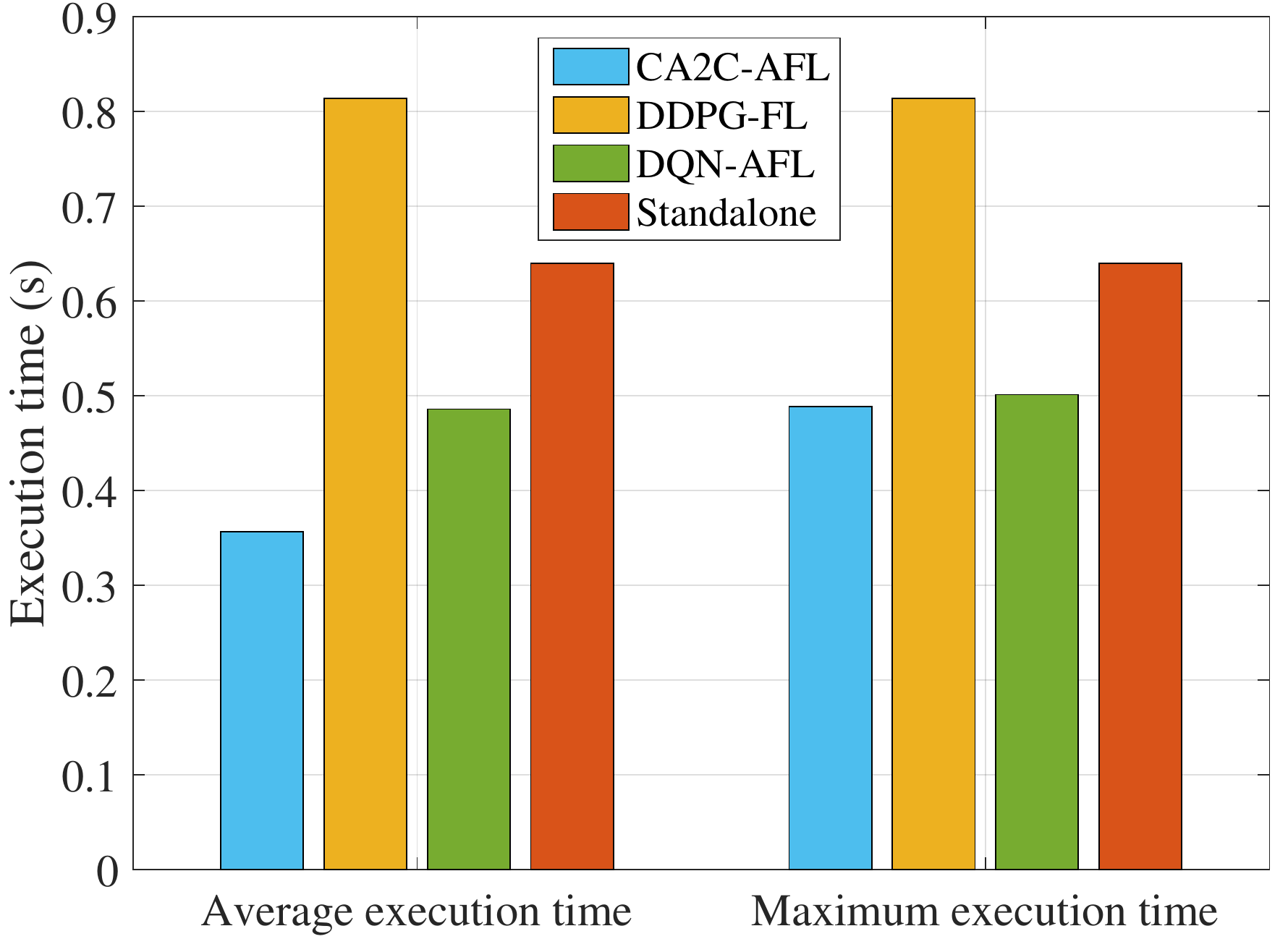}
\caption{Comparisons of average execution time and maximum execution time for anomaly detection model training.}
\end{figure}

Fig. 8 shows the average execution time and maximum execution time of the four different algorithms over the total 200 episodes. The execution time for the anomaly detection model training consists of local model update latency, local model upload latency, global model aggregation latency, and global model distribution latency within all 50 time slots. From Fig. 8, we can see that the proposed CA2C-AFL algorithm has the shortest execution time, and the DDPG-FL algorithm has the longest execution time. This is because the DDPG-FL algorithm needs to wait for all UAVs to complete their local model updates before the global aggregation process can be executed. Besides, although the average execution time and maximum execution time of the DDPG-FL algorithm and standalone algorithm are the same, the average execution time of the proposed CA2C-AFL algorithm and DQN-AFL algorithm is less than their maximum execution time due to the presence of the UAV subset selection process.

\section{Conclusion}
In this paper, we studied a VHetNet-enabled AFL-based anomaly detection framework for ubiquitous IoT devices with the assistance of a network scheduling strategy, which aimed to improve learning efficiency and detection accuracy. More specifically, the framework was designed to train local anomaly detection models at UAVs based on their sensory data and aggregated local models at a HAPS in an asynchronous manner. This aimed to decrease the federated learning execution time as well as the computation and transmission overheads. To ensure the secure transmission between UAVs and the HAPS, we adopted a differentially private WGAN-GP as the local anomaly detection model. Moreover, considering the limited onboard energy storage of UAVs, we formulated a joint device association, UAV selection, and UAV trajectory planning problem, which we solved using the CA2C approach to facilitate the efficient implementation of the self-scheduling anomaly detection model. Simulation results validated the effectiveness of the proposed framework in terms of efficiency and accuracy.

\bibliographystyle{IEEEtran}
\bibliography{IEEEabrv,reference}

\end{document}